
\input hyperbasics

\input amssym.tex 

\def\unredoffs{}
\tolerance=1000\hfuzz=2pt
\catcode`\@=11 
\ifx\hyperdef\UNd@FiNeD\def\hyperdef#1#2#3#4{#4}\def\hyperref#1#2#3#4{#4}\def\href#1#2{#2}\fi
\magnification=1200\unredoffs\baselineskip=16pt plus 2pt minus 1pt
\def\Date#1{\vfill\leftline{#1}\tenpoint\supereject%
\footline={\hss\tenrm\hyperdef\hypernoname{page}\folio\folio\hss}}%

{\count255=\time\divide\count255 by 60 \xdef\hourmin{\number\count255}
 \multiply\count255 by-60\advance\count255 by\time
 \xdef\hourmin{\hourmin:\ifnum\count255<10 0\fi\the\count255}
}
\def\date{\number\day.\number\month.\number\year\ at \hourmin}


\def\nolabels{\def\wrlabeL##1{}\def\eqlabeL##1{}\def\reflabeL##1{}}
\def\writelabels{\def\wrlabeL##1{\leavevmode\vadjust{\rlap{\smash%
{\line{{\escapechar=` \hfill\rlap{\sevenrm\hskip.03in\string##1}}}}}}}%
\def\eqlabeL##1{{\escapechar-1\rlap{\sevenrm\hskip.05in\string##1}}}%
\def\reflabeL##1{\noexpand\llap{\noexpand\sevenrm\string\string\string##1}}}
\nolabels

\global\newcount\secno \global\secno=0
\global\newcount\meqno \global\meqno=1
\def\s@csym{}

\def\newsec#1\par{\global\advance\secno by1%
{\toks0{#1}\message{(\the\secno. \the\toks0)}}%
\global\subsecno=0\eqnres@t\let\s@csym\secsym\xdef\secn@m{\the\secno}\noindent
{\bf\hyperdef\hypernoname{section}{\the\secno}{\the\secno.} #1}%
\writetoca{{\string\hyperref{}{section}{\the\secno}{\bf \the\secno\quad}} {\bf #1}}\par%
\nobreak\medskip\nobreak\noindent\ignorespaces}
\def\eqnres@t{\xdef\secsym{\the\secno.}\global\meqno=1\bigbreak\bigskip}
\def\sequentialequations{\def\eqnres@t{\bigbreak}}\xdef\secsym{}

\global\newcount\subsecno \global\subsecno=0
\def\subsec#1\par{\global\advance\subsecno by1%
{\toks0{#1}\message{(\s@csym\the\subsecno. \the\toks0)}}%
\global\subsubsecno=0%
\ifnum\lastpenalty>9000\else\bigbreak\fi
\noindent{\it\hyperdef\hypernoname{subsection}{\secn@m.\the\subsecno}%
{\secn@m.\the\subsecno.} #1}\writetoca{\string\hskip1.45cm
{\string\hyperref{}{subsection}{\secn@m.\the\subsecno}{\secn@m.\the\subsecno.}}
{#1}}\par\nobreak\medskip\nobreak\noindent\ignorespaces}

\global\newcount\subsubsecno \global\subsubsecno=0
\def\subsubsec#1\par{\global\advance\subsubsecno by1%
{\toks0{#1}\message{(\secn@m.\the\subsecno.\the\subsubsecno. \the\toks0)}}%
\global\subsubsubsecno=0%
\ifnum\lastpenalty>9000\else\bigbreak\fi
\noindent{\it\hyperdef\hypernoname{subsubsection}{\secn@m.\the\subsecno\the\subsubsecno}%
{\secn@m.\the\subsecno.\the\subsubsecno.} #1}
\par\nobreak\medskip\nobreak\noindent\ignorespaces}

\global\newcount\subsubsubsecno \global\subsubsubsecno=0
\def\subsubsubsec#1\par{\global\advance\subsubsubsecno by1%
{\toks0{#1}\message{(\secn@m.\the\subsecno.\the\subsubsecno.\the\subsubsubsecno \the\toks0)}}%
\ifnum\lastpenalty>9000\else\bigbreak\fi
\noindent{\it\hyperdef\hypernoname{subsubsection}{\secn@m.\the\subsecno\the\subsubsecno\the\subsubsubsecno}%
{\secn@m.\the\subsecno.\the\subsubsecno.\the\subsubsubsecno.} #1}%
\par\nobreak\medskip\nobreak\noindent\ignorespaces}


\def\newnewsec#1#2\par{\global\advance\secno by1%
{\toks0{#2}\message{(\secn@m. \the\toks0)}}%
\global\subsecno=0\global\subsubsecno=0\eqnres@t\let\s@csym\secsym\xdef\secn@m{\the\secno}\noindent
\ifnum\lastpenalty>9000\else\bigbreak\fi
\noindent{\bf\hyperdef\hypernoname{section}{\secn@m}{\secn@m.} #2}%
\writetoca{{\string\hyperref{}{section}{\the\secno}{\bf \the\secno\quad}} {\bf #2}}
\DefWarn#1%
\xdef#1{\noexpand\hyperref{}{section}{\the\secno}%
{\the\secno}}\writedef{#1\leftbracket#1}\wrlabeL{#1=#1}%
\par\nobreak\medskip\nobreak\noindent\ignorespaces}

\def\newsubsec#1#2\par{\global\advance\subsecno by1%
{\toks0{#2}\message{(\secn@m.\the\subsecno. \the\toks0)}}%
\global\subsubsecno=0%
\ifnum\lastpenalty>9000\else\bigbreak\fi
\noindent{\it\hyperdef\hypernoname{subsection}{\secn@m.\the\subsecno}%
{\secn@m.\the\subsecno.} #2}
\DefWarn#1%
\xdef#1{\noexpand\hyperref{}{subsection}{\secn@m.\the\subsecno}%
{\secn@m.\the\subsecno}}\writedef{#1\leftbracket#1}\wrlabeL{#1=#1}%
\writetoca{\string\hskip1.45cm
{\string\hyperref{}{subsection}{\secn@m.\the\subsecno}{\secn@m.\the\subsecno.}}
{#2}}%
\par\nobreak\medskip\nobreak\noindent\ignorespaces}

\def\newsubsecstar#1#2\par{\global\advance\subsecno by1%
{\toks0{#2}\message{(\secn@m.\the\subsecno. \the\toks0)}}%
\global\subsubsecno=0%
\ifnum\lastpenalty>9000\else\bigbreak\fi
\noindent{\it\hyperdef\hypernoname{subsection}{\secn@m.\the\subsecno}%
{\secn@m.\the\subsecno.} #2}
\DefWarn#1%
\xdef#1{\noexpand\hyperref{}{subsection}{\secn@m.\the\subsecno}%
{\secn@m.\the\subsecno}}\writedef{#1\leftbracket#1}\wrlabeL{#1=#1}%
\par\nobreak\medskip\nobreak\noindent\ignorespaces}

\def\newsubsubsec#1#2\par{\global\advance\subsubsecno by1%
{\toks0{#2}\message{(\secn@m.\the\subsecno.\the\subsubsecno. \the\toks0)}}%
\global\subsubsubsecno=0%
\ifnum\lastpenalty>9000\else\bigbreak\fi
\noindent{\it\hyperdef\hypernoname{subsubsection}{\secn@m.\the\subsecno.\the\subsubsecno}%
{\secn@m.\the\subsecno.\the\subsubsecno.} #2}
\DefWarn#1%
\xdef#1{\noexpand\hyperref{}{subsubsection}{\secn@m.\the\subsecno.\the\subsubsecno}%
{\secn@m.\the\subsecno.\the\subsubsecno}}\writedef{#1\leftbracket#1}\wrlabeL{#1=#1}%
\par\nobreak\medskip\nobreak\noindent\ignorespaces}

\def\newsubsubsubsec#1#2\par{\global\advance\subsubsubsecno by1%
{\toks0{#2}\message{(\secn@m.\the\subsecno.\the\subsubsecno.\the\subsubsubsecno \the\toks0)}}%
\ifnum\lastpenalty>9000\else\bigbreak\fi
\noindent{\it\hyperdef\hypernoname{subsubsection}{\secn@m.\the\subsecno\the\subsubsecno\the\subsubsubsecno}%
{\secn@m.\the\subsecno.\the\subsubsecno.\the\subsubsubsecno.} #2}
\DefWarn#1%
\xdef#1{\noexpand\hyperref{}{subsubsubsection}{\secn@m.\the\subsecno.\the\subsubsecno.\the\subsubsubsecno}%
{\secn@m.\the\subsecno.\the\subsubsecno.\the\subsubsubsecno}}\writedef{#1\leftbracket#1}\wrlabeL{#1=#1}%
\par\nobreak\medskip\nobreak\noindent\ignorespaces}

\def\appendix#1#2{\global\meqno=1\global\subsecno=0\global\subsubsecno=0\xdef\secsym{\hbox{#1.}}%
\bigbreak\bigskip\noindent{\bf Appendix \hyperdef\hypernoname{appendix}{#1}%
{#1.} #2}{\toks0{(#1. #2)}\message{\the\toks0}}%
\xdef\s@csym{#1.}\xdef\secn@m{#1}%
\writetoca{{\string\hyperref{}{appendix}{#1}{\bf {#1}\quad}} {\bf #2}}%
\par\nobreak\medskip\nobreak}

%
\def\checkm@de#1#2{\ifmmode{\def\f@rst##1{##1}\hyperdef\hypernoname{equation}%
{#1}{#2}}\else\hyperref{}{equation}{#1}{#2}\fi}
\def\eqnn#1{\DefWarn#1\xdef #1{(\noexpand\relax\noexpand\checkm@de%
{\s@csym\the\meqno}{\secsym\the\meqno})}%
\wrlabeL#1\writedef{#1\leftbracket#1}\global\advance\meqno by1}
\def\f@rst#1{\c@t#1a\em@ark}\def\c@t#1#2\em@ark{#1}
\def\eqna#1{\DefWarn#1\wrlabeL{#1$\{\}$}%
\xdef #1##1{(\noexpand\relax\noexpand\checkm@de%
{\s@csym\the\meqno\noexpand\f@rst{##1}1}{\hbox{$\secsym\the\meqno##1$}})}
\writedef{#1\numbersign1\leftbracket#1{\numbersign1}}\global\advance\meqno by1}
\def\eqn#1#2{\DefWarn#1%
\xdef #1{(\noexpand\hyperref{}{equation}{\s@csym\the\meqno}%
{\secsym\the\meqno})}$$#2\eqno(\hyperdef\hypernoname{equation}%
{\s@csym\the\meqno}{\secsym\the\meqno})\eqlabeL#1$$%
\writedef{#1\leftbracket#1}\global\advance\meqno by1}
\def\xeqn{\expandafter\xe@n}\def\xe@n(#1){#1}
\def\xeqna#1{\expandafter\xe@n#1}
\def\eqns#1{(\e@ns #1{\hbox{}})}
\def\e@ns#1{\ifx\UNd@FiNeD#1\message{eqnlabel \string#1 is undefined.}%
\xdef#1{(?.?)}\fi{\let\hyperref=\relax\xdef\next{#1}}%
\ifx\next\em@rk\def\next{}\else%
\ifx\next#1\xeqn#1\else\def\n@xt{#1}\ifx\n@xt\next#1\else\xeqna#1\fi
\fi\let\next=\e@ns\fi\next}
\def\DefWarn#1{}
%
\newskip\footskip\footskip14pt plus 1pt minus 1pt 
\def\footnotefont{\ninepoint}\def\f@t#1{\footnotefont #1\@foot}
\def\f@@t{\baselineskip\footskip\bgroup\footnotefont\aftergroup\@foot\let\next}
\setbox\strutbox=\hbox{\vrule height9.5pt depth4.5pt width0pt}
\global\newcount\ftno \global\ftno=0
\def\foot{\global\advance\ftno by1\def\foot@rg{\hyperref{}{footnote}%
{\the\ftno}{\the\ftno}\xdef\foot@rg{\noexpand\hyperdef\noexpand\hypernoname%
{footnote}{\the\ftno}{\the\ftno}}}\footnote{$^{\foot@rg}$}}
%
%
%
\global\newcount\refno \global\refno=1
\newwrite\rfile
\def\ref{[\hyperref{}{reference}{\the\refno}{\the\refno}]\nref}
\def\nref#1{\DefWarn#1%
\xdef#1{[\noexpand\hyperref{}{reference}{\the\refno}{\the\refno}]}%
\writedef{#1\leftbracket#1}%
\ifnum\refno=1\immediate\openout\rfile=\jobname.refs\fi
\chardef\wfile=\rfile\immediate\write\rfile{\noexpand\item{[\noexpand\hyperdef%
\noexpand\hypernoname{reference}{\the\refno}{\the\refno}]\ }%
\reflabeL{#1\hskip.31in}\pctsign}\global\advance\refno by1\findarg}
\def\findarg#1#{\begingroup\obeylines\newlinechar=`\^^M\pass@rg}
{\obeylines\gdef\pass@rg#1{\writ@line\relax #1^^M\hbox{}^^M}%
\gdef\writ@line#1^^M{\expandafter\toks0\expandafter{\striprel@x #1}%
\edef\next{\the\toks0}\ifx\next\em@rk\let\next=\endgroup\else\ifx\next\empty%
\else\immediate\write\wfile{\the\toks0}\fi\let\next=\writ@line\fi\next\relax}}
\def\striprel@x#1{} \def\em@rk{\hbox{}}
\def\lref{\begingroup\obeylines\lr@f}
\def\lr@f#1#2{\DefWarn#1\gdef#1{\let#1=\UNd@FiNeD\ref#1{#2}}\endgroup\unskip}
\def\semi{;\hfil\break}
\def\addref#1{\immediate\write\rfile{\noexpand\item{}#1}} 
\def\listrefs{\vfill\supereject\immediate\closeout\rfile\writestoppt
\baselineskip=\footskip\centerline{{\bf References}}\bigskip{\parindent=20pt%
\frenchspacing\escapechar=` \input \jobname.refs\vfill\eject}\nonfrenchspacing}
\def\startrefs#1{\immediate\openout\rfile=\jobname.refs\refno=#1}
\def\xref{\expandafter\xr@f}\def\xr@f[#1]{#1}
\def\refs#1{\count255=1[\r@fs #1{\hbox{}}]}
\def\r@fs#1{\ifx\UNd@FiNeD#1\message{reflabel \string#1 is undefined.}%
\nref#1{need to supply reference \string#1.}\fi%
\vphantom{\hphantom{#1}}{\let\hyperref=\relax\xdef\next{#1}}%
\ifx\next\em@rk\def\next{}%
\else\ifx\next#1\ifodd\count255\relax\xref#1\count255=0\fi%
\else#1\count255=1\fi\let\next=\r@fs\fi\next}
%

%
\newwrite\ffile\global\newcount\figno \global\figno=1
\def\fig{fig.~\hyperref{}{figure}{\the\figno}{\the\figno}\nfig}
\def\nfig#1{\DefWarn#1%
\xdef#1{fig.~\noexpand\hyperref{}{figure}{\the\figno}{\the\figno}}%
\writedef{#1\leftbracket fig.\noexpand~\xfig#1}%
\ifnum\figno=1\immediate\openout\ffile=\jobname.figs\fi\chardef\wfile=\ffile%
{\let\hyperref=\relax
\immediate\write\ffile{\noexpand\medskip\noexpand\item{Fig.\ %
\noexpand\hyperdef\noexpand\hypernoname{figure}{\the\figno}{\the\figno}. }
\reflabeL{#1\hskip.55in}\pctsign}}\global\advance\figno by1\findarg}
\def\xfig{\expandafter\xf@g}\def\xf@g fig.\penalty\@M\ {}
\def\figs#1{figs.~\f@gs #1{\hbox{}}}
\def\f@gs#1{{\let\hyperref=\relax\xdef\next{#1}}\ifx\next\em@rk\def\next{}\else
\ifx\next#1\xfig #1\else#1\fi\let\next=\f@gs\fi\next}
%
\def\figin{\epsfcheck\figin}\def\figins{\epsfcheck\figins}
\def\epsfcheck{\ifx\epsfbox\UnDeFiNeD
\message{(NO epsf.tex, FIGURES WILL BE IGNORED)}
\gdef\figin##1{\vskip2in}\gdef\figins##1{\hskip.5in}
\else\message{(FIGURES WILL BE INCLUDED)}%
\gdef\figin##1{##1}\gdef\figins##1{##1}\fi}
\def\figinsert{\goodbreak\topinsert}
\def\ifig#1#2#3{\DefWarn#1\xdef#1{fig.~\the\figno}
\writedef{#1\leftbracket fig.\noexpand~\the\figno}%
\figinsert\figin{\centerline{#3}}
\smallskip
\leftskip=0pt \rightskip=0pt
\baselineskip12pt\noindent
{{\bf Fig.~\the\figno}\ \ninepoint #2}
\medskip
\global\advance\figno by1\par\endinsert}
\newwrite\lfile
{\escapechar-1\xdef\pctsign{\string\%}\xdef\leftbracket{\string\{}
\xdef\rightbracket{\string\}}\xdef\numbersign{\string\#}}
\def\writedefs{\immediate\openout\lfile=label.defs \def\writedef##1{%
{\let\hyperref=\relax\let\hyperdef=\relax\let\hypernoname=\relax
 \immediate\write\lfile{\string\checkdef\string##1\rightbracket}}}}%
\def\writestop{\def\writestoppt{\immediate\write\lfile{\string\pageno
 \the\pageno\string\startrefs\leftbracket\the\refno\rightbracket
 \string\def\string\secsym\leftbracket\secsym\rightbracket
 \string\secno\the\secno\string\meqno\the\meqno}\immediate\closeout\lfile}}
\def\writestoppt{}\def\writedef#1{}

\def\seclab#1\par{\DefWarn#1%
\xdef #1{\noexpand\hyperref{}{section}{\the\secno}{\the\secno}}%
\writedef{#1\leftbracket#1}\wrlabeL{#1=#1}\par%
\nobreak\medskip\nobreak\noindent\ignorespaces}
\def\subseclab#1\par{\DefWarn#1%
\xdef #1{\noexpand\hyperref{}{subsection}{\the\secno.\the\subsecno}%
{\the\secno.\the\subsecno}}\writedef{#1\leftbracket#1}\wrlabeL{#1=#1}\par%
\nobreak\medskip\nobreak\noindent\ignorespaces}
\def\subsubseclab#1\par{\DefWarn#1%
\xdef#1{\noexpand\hyperref{}{subsubsection}{\the\secno.\the\subsecno.\the\subsubsecno}%
{\the\secno.\the\subsecno.\the\subsubsecno}}\writedef{#1\leftbracket#1}\wrlabeL{#1=#1}\par%
\nobreak\medskip\nobreak\noindent\ignorespaces}
\def\applab#1\par{\DefWarn#1%
\xdef#1{\noexpand\hyperref{}{appendix}{\secn@m}{\secn@m}}%
\writedef{#1\leftbracket#1}\wrlabeL{#1=#1}%
\par\nobreak\medskip\nobreak\noindent\ignorespaces}
\def\appsublab#1{\DefWarn#1%
\xdef #1{\noexpand\hyperref{}{appendix}{\secn@m.\the\subsecno}{\secn@m.\the\subsecno}}%
\writedef{#1\leftbracket#1}\wrlabeL{#1=#1}}
\newwrite\tfile \def\writetoca#1{}
\def\leaderfill{\leaders\hbox to 1em{\hss.\hss}\hfill}
\def\writetoc{\immediate\openout\tfile=\jobname.toc
   \def\writetoca##1{{\edef\next{\write\tfile{\noindent ##1
   \string\leaderfill{
   \string\hyperref{}{page}{\noexpand\number\pageno}%
   {\noexpand\number\pageno}} \par}}\next}}
}
\newread\ch@ckfile
\def\listtoc{\immediate\closeout\tfile\immediate\openin\ch@ckfile=\jobname.toc
\ifeof\ch@ckfile\message{no file \jobname.toc, no table of contents this pass}%
\else\closein\ch@ckfile\centerline{\bf Contents}\nobreak\medskip%
{\baselineskip=15.5pt\footnotefont\parskip=0pt\catcode`\@=11\input\jobname.toc
\catcode`\@=12\bigbreak\bigskip}\fi}
\catcode`\@=12 
\def\tenpoint{\def\rm{\fam0\tenrm}
\textfont0=\tenrm \scriptfont0=\sevenrm \scriptscriptfont0=\fiverm
\textfont1=\teni  \scriptfont1=\seveni  \scriptscriptfont1=\fivei
\textfont2=\tensy \scriptfont2=\sevensy \scriptscriptfont2=\fivesy
\textfont\itfam=\tenit \def\it{\fam\itfam\tenit}\def\footnotefont{\ninepoint}%
\textfont\bffam=\tenbf \def\bf{\fam\bffam\tenbf}\def\sl{\fam\slfam\tensl}\rm}
\font\ninerm=cmr9 \font\sixrm=cmr6 \font\ninei=cmmi9 \font\sixi=cmmi6
\font\ninesy=cmsy9 \font\sixsy=cmsy6 \font\ninebf=cmbx9
\font\nineit=cmti9 \font\ninesl=cmsl9 \skewchar\ninei='177
\skewchar\sixi='177 \skewchar\ninesy='60 \skewchar\sixsy='60
\def\ninepoint{\def\rm{\fam0\ninerm}
\textfont0=\ninerm \scriptfont0=\sixrm \scriptscriptfont0=\fiverm
\textfont1=\ninei \scriptfont1=\sixi \scriptscriptfont1=\fivei
\textfont2=\ninesy \scriptfont2=\sixsy \scriptscriptfont2=\fivesy
\textfont\itfam=\ninei \def\it{\fam\itfam\nineit}\def\sl{\fam\slfam\ninesl}%
\textfont\bffam=\ninebf \def\bf{\fam\bffam\ninebf}\rm}
%
\hyphenation{anom-aly anom-alies coun-ter-term coun-ter-terms}

\def\tikzcaption#1#2{\DefWarn#1\xdef#1{Fig.~\the\figno}
\writedef{#1\leftbracket Fig.\noexpand~\the\figno}%
{
\smallskip
\leftskip=20pt \rightskip=20pt \baselineskip12pt\noindent
{{\bf Fig.~\the\figno}\ \ninepoint #2}
\bigskip
\global\advance\figno by1 \par}}

\def\ntoalpha#1{%
\ifcase#1%
@%
\or A\or B\or C\or D\or E\or F\or G\or H\or I\or J\or K\or L\or M%
\fi
}

\global\newcount\appno \global\appno=1
\def\applab#1{\xdef #1{\ntoalpha{\appno}}\writedef{#1\leftbracket#1}\wrlabeL{#1=#1}
\global\advance\appno by1}

\def\preprint#1 #2\par{\rightline{\vbox{\baselineskip12pt\hbox{#1}\hbox{#2}}}\vskip2cm}
%
\def\title#1\par{\centerline{\bf #1}\nopagenumbers\pageno=0}
\def\author#1\par{\bigskip\bigskip\centerline{#1}}

\newcount\addressno

\def\email#1#2{
\footnote{\null}{\kern-\parindent \llap{$^#1$\hskip1pt}email: #2}}

\def\startcenter{%
  \par
  \begingroup
  \leftskip=0pt plus 1fil
  \rightskip=\leftskip
  \parindent=0pt
  \parfillskip=0pt
}
\def\stopcenter{\endgroup}

\def\address{\bigskip%
  \ifnum\the\addressno=0\else\stopcenter\endgroup\fi
  \advance\addressno by 1%
  \begingroup
  \startcenter
  \it
  \obeylines
  \addressAux
}
\def\addressAux#1{#1}

\def\abstract{\stopcenter\endgroup\bigskip\bigskip\noindent}

\def\Dsl{\,\raise.15ex\hbox{/}\mkern-13.5mu D} 
\def\dsl{\raise.15ex\hbox{/}\kern-.57em\partial}
 
\def\boxeqn#1{\vcenter{\vbox{\hrule\hbox{\vrule\kern3pt\vbox{\kern3pt
	\hbox{${\displaystyle #1}$}\kern3pt}\kern3pt\vrule}\hrule}}}


\def\ap{{\alpha^{\prime}}}

\def\a{\alpha}
\def\b{{\beta}}
\def\g{{\gamma}}
\def\d{{\delta}}

\def\l{\lambda}

\def\s{{\sigma}}
\def\t{{\theta}}

\def\half{{1\over 2}}
\def\p{{\partial}}

\def\({\left(}
\def\){\right)}

\def\cW{{\cal W}}


\def\AYM{A^{\rm SYM}}



\def\sfrac#1/#2{\kern.1em\raise.5ex\hbox{\the\scriptfont0 #1}%
\kern-.1em/\kern-.15em\lower.25ex\hbox{\the\scriptfont0 #2}}



\def\qed{\hbox{\hskip 3pt
\vbox{\hrule\hbox to 7pt{\vrule height 7pt\hfill\vrule}
\hrule}}\hskip3pt}

\overfullrule=0pt\relax

\frenchspacing

\def\checkdef#1#2{
\ifx\UndeFined#1%
	\def#1{#2}
\else
	\immediate\write16{*** BUG ***: the label \string#1 is already defined ***}
\fi
}
\newread\instream

\openin\instream= label.defs
\ifeof\instream\message{No labels in advance yet. Wait till next pass.}
\else\closein\instream \input label.defs
\fi
\writedefs

\def\arXiv:#1].{\hepthStrip#1 \nil}
\def\hepthStrip#1 #2\nil{\href{http://arxiv.org/abs/#1}{arXiv:#1 #2\unskip}].}


\input amssym
\input epsf

\def\textbf#1{{\bf #1}}
\def\paragraph#1{\medskip\noindent{\it #1.}}
\def\frac#1#2{{#1\over#2}}
\def\AFq(#1,#2,#3,#4,#5){A^{F^4}_{#1#2#3#4#5}}
\def\res#1{\mathop{\rm Res}_{#1}}


\title Massless representation of massive superfields and tree amplitudes

\title with the pure spinor formalism

\author
Sitender Pratap Kashyap\email{\star}{sitender@cmi.ac.in}$^\star$, Carlos R. Mafra\email{\ap}{c.r.mafra@soton.ac.uk}$^\ap$,
Mritunjay Verma\email{\dagger}{mritunjay@iiti.ac.in}$^\dagger$, Luis
Ypanaqu\'e\email{\bullet}{luis.12yr@gmail.com}$^\bullet$

\address
$^\star$ Chennai Mathematical Institute, H1 SIPCOT IT Park,
Kelambakkam, Tamil Nadu, India 603103

$^\ap$Mathematical Sciences and STAG Research Centre, University of Southampton,
Highfield, Southampton, SO17 1BJ, UK

$^\dagger$ Indian Institute of Technology Indore, Khandwa Road,
Simrol, Indore 453552, India

\abstract
We construct the unintegrated vertex operator at the first mass level of the open superstring
from the OPE of massless vertices.
Using BRST cohomology
manipulations, the tree amplitude of two massless and one massive
state is rewritten in terms of the pure spinor superspace kinematic expression of the massless four-point
amplitude at the $\ap^2$ level. A generalization relating the partial $n$-point tree amplitudes
with one massive state and linear combinations of the $\ap^2$ corrections to $n{+}1$ massless
amplitudes is found and shown to be consistent with unitarity.

\Date{July 2024}



\lref\massivevone{
	S.P.~Kashyap, C.R.~Mafra, M.~Verma and L.A.~Ypanaque,
	``A relation between massive and massless string tree amplitudes,''
	[arXiv:2311.12100v1 [hep-th]].
}
\lref\nptMethod{
	C.~R.~Mafra, O.~Schlotterer, S.~Stieberger and D.~Tsimpis,
	``A recursive method for SYM n-point tree amplitudes,''
	Phys.\ Rev.\ D {\bf 83}, 126012 (2011).
	[arXiv:1012.3981 [hep-th]].
}
\lref\GSteight{
	M.B.~Green and J.H.~Schwarz,
	``Supersymmetrical Dual String Theory. 2. Vertices and Trees,''
	Nucl. Phys. B \textbf{198}, 252-268 (1982)
}
\lref\PSteight{
	C.~R.~Mafra,
	``Four-point one-loop amplitude computation in the pure spinor formalism,''
	JHEP \textbf{01}, 075 (2006)
	[arXiv:hep-th/0512052 [hep-th]].
}
\lref\towards{
	C.R.~Mafra,
	``Towards Field Theory Amplitudes From the Cohomology of Pure Spinor Superspace,''
	JHEP {\bf 1011}, 096 (2010).
	[arXiv:1007.3639 [hep-th]].
}
\lref\towardsmass{
	C.~R.~Mafra,
	``Towards massive field-theory amplitudes from the cohomology of pure spinor superspace,''
	JHEP \textbf{11}, 045 (2024)
	[arXiv:2407.11849 [hep-th]].
}
\lref\sixtree{
	C.R.~Mafra, O.~Schlotterer, S.~Stieberger and D.~Tsimpis,
	``Six Open String Disk Amplitude in Pure Spinor Superspace,''
	Nucl. Phys. B \textbf{846}, 359-393 (2011)
	[arXiv:1011.0994 [hep-th]].
}
\lref\Soares{
	B.~R.~Soares,
	``Constructing massive superstring vertex operators from massless vertex operators using the pure spinor formalism,''
	Phys. Lett. B \textbf{852}, 138611 (2024)
	[arXiv:2401.03208 [hep-th]].
}

\lref\bianchifac{
	M.~Bianchi and A.~L.~Guerrieri,
	``On the soft limit of open string disk amplitudes with massive states,''
	JHEP \textbf{09}, 164 (2015)
	[arXiv:1505.05854 [hep-th]].
}

\lref\website{
	http://www.southampton.ac.uk/\~{}crm1n16/pss.html
}
\lref\EOMbbs{
	C.R.~Mafra and O.~Schlotterer,
  	``Multiparticle SYM equations of motion and pure spinor BRST blocks,''
	JHEP {\bf 1407}, 153 (2014).
	[arXiv:1404.4986 [hep-th]].
}
\lref\BGap{
	C.R.~Mafra and O.~Schlotterer,
  	``Non-abelian $Z$-theory: Berends-Giele recursion for the $\alpha'$-expansion of disk integrals,''
	[arXiv:1609.07078].
	{\tt http://repo.or.cz/BGap.git}
}

\lref\partIcohomology{
	C.R.~Mafra and O.~Schlotterer,
	``Cohomology foundations of one-loop amplitudes in pure spinor superspace,''
	[arXiv:1408.3605 [hep-th]].
}

\lref\cdescent{
	C.R.~Mafra,
	``KK-like relations of $\alpha$' corrections to disk amplitudes,''
	JHEP \textbf{03}, 012 (2022)
	[arXiv:2108.01081 [hep-th]].
}

\lref\massSweden{
       	M.~Guillen, H.~Johansson, R.~L.~Jusinskas and O.~Schlotterer,
	``Scattering Massive String Resonances through Field-Theory Methods,''
	Phys. Rev. Lett. \textbf{127}, no.5, 051601 (2021)
	[arXiv:2104.03314 [hep-th]].
}
\lref\oneloopbb{
	C.R.~Mafra and O.~Schlotterer,
	``The Structure of n-Point One-Loop Open Superstring Amplitudes,''
	JHEP \textbf{08}, 099 (2014)
	[arXiv:1203.6215 [hep-th]].
}

\lref\fivetree{
	C.R.~Mafra,
	``Simplifying the Tree-level Superstring Massless Five-point Amplitude,''
	JHEP {\bf 1001}, 007 (2010).
	[arXiv:0909.5206 [hep-th]].
}
\lref\nptTree{
	C.R.~Mafra, O.~Schlotterer and S.~Stieberger,
	``Complete N-Point Superstring Disk Amplitude I. Pure Spinor Computation,''
	Nucl.\ Phys.\ B {\bf 873}, 419 (2013).
	[arXiv:1106.2645 [hep-th]].
}
\lref\nptTreeII{
	C.~R.~Mafra, O.~Schlotterer and S.~Stieberger,
	``Complete N-Point Superstring Disk Amplitude II. Amplitude
	and Hypergeometric Function Structure,''
	Nucl.\ Phys.\ B {\bf 873}, 461 (2013).
	[arXiv:1106.2646 [hep-th]].
}
\lref\fourtree{
	C.R.~Mafra,
	``Pure Spinor Superspace Identities for Massless Four-point Kinematic Factors,''
	JHEP \textbf{04}, 093 (2008)
	[arXiv:0801.0580 [hep-th]].
}
\lref\PSthreemass{
	S.~Chakrabarti, S.~P.~Kashyap and M.~Verma,
	``Amplitudes Involving Massive States Using Pure Spinor Formalism,''
	JHEP \textbf{12}, 071 (2018)
	[arXiv:1808.08735 [hep-th]].
}
\lref\drinfeld{
	J.~Broedel, O.~Schlotterer, S.~Stieberger and T.~Terasoma,
  	``All order $\alpha^{\prime}$-expansion of superstring trees from the Drinfeld associator,''
	Phys.\ Rev.\ D {\bf 89}, no. 6, 066014 (2014).
	[arXiv:1304.7304 [hep-th]].
}

\lref\FMS{
	D.~Friedan, E.J.~Martinec and S.H.~Shenker,
  	``Conformal Invariance, Supersymmetry and String Theory,''
  	Nucl.\ Phys.\ B {\bf 271} (1986) 93.
}

\lref\Chakrabarti{
	S.~Chakrabarti, S.~P.~Kashyap and M.~Verma,
	``Integrated Massive Vertex Operator in Pure Spinor Formalism,''
	JHEP \textbf{10}, 147 (2018)
	[arXiv:1802.04486 [hep-th]].
}
\lref\masstheta{
	S.~Chakrabarti, S.~P.~Kashyap and M.~Verma,
	``Theta Expansion of First Massive Vertex Operator in Pure Spinor,''
	JHEP \textbf{01}, 019 (2018)
	[arXiv:1706.01196 [hep-th]].
}

\lref\Koh{
	I.~G.~Koh, W.~Troost and A.~Van Proeyen,
	``Covariant Higher Spin Vertex Operators in the Ramond Sector,''
	Nucl. Phys. B \textbf{292}, 201-221 (1987)
}

\lref\ICTP{
	N.~Berkovits,
  	``ICTP lectures on covariant quantization of the superstring,''
	[hep-th/0209059].
}
\lref\axioms{
	A.~Matsuo and K.~Nagatomo,
	``On axioms for a vertex algebra and the locality of quantum fields,''
	[arXiv:hep-th/9706118 [hep-th]].
}

\lref\japwick{
	T.~Takagi and T.~Yoshikawa,
	``Generalized Wick theorems in conformal field theory and the Borcherds identity,''
	J. Phys. Soc. Jap. \textbf{87}, no.11, 114007 (2018)
	[arXiv:1604.04032 [math-ph]].
}
\lref\borcherds{
	R.~E.~Borcherds,
	``Vertex algebras, Kac-Moody algebras, and the monster,''
	Proc. Nat. Acad. Sci. \textbf{83}, 3068-3071 (1986)
}

\lref\farril{
	J.~M.~Figueroa-O'Farrill,
	``N=2 structures in all string theories,''
	J. Math. Phys. \textbf{38}, 5559-5575 (1997)
	[arXiv:hep-th/9507145 [hep-th]].
}
\lref\thielemans{
	K.~Thielemans,
	``An Algorithmic approach to operator product expansions, W algebras and W strings,''
	[arXiv:hep-th/9506159 [hep-th]].
}
\lref\BCpaper{
	N.~Berkovits and O.~Chandia,
	``Massive superstring vertex operator in D = 10 superspace,''
	JHEP \textbf{08}, 040 (2002)
	[arXiv:hep-th/0204121 [hep-th]].
}
\lref\yellowbook{
	P.~Di Francesco, P.~Mathieu and D.~Senechal,
	``Conformal Field Theory,''
	Springer-Verlag, 1997,
	ISBN 978-0-387-94785-3, 978-1-4612-7475-9
	doi:10.1007/978-1-4612-2256-9
}
\lref\bais{
	F.~A.~Bais, P.~Bouwknegt, M.~Surridge and K.~Schoutens,
	``Extensions of the Virasoro Algebra Constructed from Kac-Moody Algebras Using Higher Order Casimir Invariants,''
	Nucl. Phys. B \textbf{304}, 348-370 (1988)
}
\lref\psf{
 	N.~Berkovits,
	``Super-Poincare covariant quantization of the superstring,''
	JHEP {\bf 0004}, 018 (2000)
	[arXiv:hep-th/0001035].
	\semi
	N.~Berkovits,
	``Pure spinor formalism as an N = 2 topological string,''
	JHEP {\bf 0510}, 089 (2005)
	[arXiv:hep-th/0509120].
}

\lref\LiE{
	M.A.A. van Leeuwen, A.M. Cohen and B. Lisser,
	``LiE, A Package for Lie Group Computations'', Computer Algebra Nederland, Amsterdam, ISBN 90-74116-02-7, 1992
}
\lref\wittentwistor{
	E.Witten,
        ``Twistor-Like Transform In Ten-Dimensions''
        Nucl.Phys. B {\bf 266}, 245~(1986)
}
\lref\higherSYM{
	C.R.~Mafra and O.~Schlotterer,
	``A solution to the non-linear equations of D=10 super Yang-Mills theory,''
	Phys.\ Rev.\ D {\bf 92}, no. 6, 066001 (2015).
	[arXiv:1501.05562 [hep-th]].
}
\lref\treereview{
	C.~R.~Mafra and O.~Schlotterer,
	``Tree-level amplitudes from the pure spinor superstring,''
	Phys. Rept. \textbf{1020}, 1-162 (2023)
	[arXiv:2210.14241 [hep-th]].
}
\lref\BGBCJ{
	C.R.~Mafra and O.~Schlotterer,
  	``Berends-Giele recursions and the BCJ duality in superspace and components,''
	JHEP {\bf 1603}, 097 (2016).
	[arXiv:1510.08846 [hep-th]].
}
\lref\FORM{
	J.A.M.~Vermaseren,
	``New features of FORM,''
	arXiv:math-ph/0010025.
\semi
	M.~Tentyukov and J.A.M.~Vermaseren,
	``The multithreaded version of FORM,''
	arXiv:hep-ph/0702279.
}
\lref\MPS{
	N.~Berkovits,
  	``Multiloop amplitudes and vanishing theorems using the pure spinor formalism for the superstring,''
	JHEP {\bf 0409}, 047 (2004).
	[hep-th/0406055].
}
\lref\facrevisited{
	Z.~Xiao and C.~J.~Zhu,
	``Factorization of the two loop four-particle amplitude in superstring theory revisited,''
	JHEP \textbf{06}, 002 (2005)
	[arXiv:hep-th/0412018 [hep-th]].
}
\lref\nimaLorenzYuSebas{
	N.~Arkani-Hamed, L.~Eberhardt, Y.~t.~Huang and S.~Mizera,
	``On unitarity of tree-level string amplitudes,''
	JHEP \textbf{02}, 197 (2022)
	[arXiv:2201.11575 [hep-th]].
}
\lref\olihigh{
	O.~Schlotterer,
	``Higher Spin Scattering in Superstring Theory,''
	Nucl. Phys. B \textbf{849}, 433-460 (2011)
	[arXiv:1011.1235 [hep-th]].
}
\lref\stielust{
	W.~Z.~Feng, D.~Lust, O.~Schlotterer, S.~Stieberger and T.~R.~Taylor,
	``Direct Production of Lightest Regge Resonances,''
	Nucl. Phys. B \textbf{843}, 570-601 (2011)
	[arXiv:1007.5254 [hep-th]].
}

\lref\PSspace{
	N.~Berkovits,
	``Explaining Pure Spinor Superspace,''
	[arXiv:hep-th/0612021 [hep-th]].
}
\lref\mackey{
	L.~Solomon, ``A Mackey formula in the group ring of a Coxeter group,''
	Journal of Algebra, 41(2), (1976) 255-264.
}

\lref\garsiaReutenauer{
       A.M.~Garsia, C.~Reutenauer, ``A decomposition of Solomon's descent algebra''.
       Advances in Mathematics, 77(2) (1989), 189-262
}
\lref\nunez{
	G.~Aldazabal, M.~Bonini and C.~A.~Nunez,
	``Covariant superstring fermionic amplitudes. vertex operators and picture changing,''
	Nucl. Phys. B \textbf{319}, 342-366 (1989)
}
\lref\refo{
	A.~Hanany, D.~Forcella and J.~Troost,
	``The Covariant perturbative string spectrum,''
	Nucl. Phys. B \textbf{846}, 212-225 (2011)
	[arXiv:1007.2622 [hep-th]].
}
\lref\reft{
	T.~Basile and C.~Markou,
	``On the deep superstring spectrum,''
	JHEP \textbf{07}, 184 (2024)
	[arXiv:2405.18467 [hep-th]].
}
\listtoc
\writetoc

\newsec{Introduction}

The successful calculation of the general massless open string tree amplitudes
\refs{\nptTree,\nptTreeII} with the pure spinor formalism \psf\ 
still does not have a counterpart involving massive states. One of the
reasons for this situation
is the added complexity in the description of massive superfields in ten dimensions and their
use in constructing massive vertex operators.

However, the pure spinor formalism comes equipped with a powerful notion of a cohomological pure spinor
superspace \PSspace. At the massless level, BRST cohomology manipulations give rise to many simplifications
that have been exploited in several papers \refs{\fourtree\fivetree\towards\sixtree{--}\nptMethod}
culminating in the $n$-point tree amplitude of \nptTree, see the review \treereview.
We wish to transfer some of the techniques and knowledge accumulated with pure spinor superspace
expressions involving massless SYM superfields into the manipulation and simplification of massive
amplitudes, with the hopes of advancing the knowledge of massive amplitudes beyond its current
limited state. The unintegrated vertex operator at the first massive level and its superfield description was
found in 2002 by Berkovits and Chandia \BCpaper, but it took many years until it was used in the
calculation of the three point amplitude with two massless and one massive state \PSthreemass.

In this paper, two major advantages of the pure spinor formalism in calculating massless tree
amplitudes will start to be transferred to
the study of massive amplitudes: the simplicity of the SYM superfield massless description and
the BRST cohomology manipulations in pure spinor superspace. To accomplish the first goal we will
construct the massive superfields for the unintegrated vertex operator
using the OPEs between massless vertex operators\foot{This construction with pure spinors was
firstly announced in the companion of this paper \massivevone\ and later confirmed by similar
calculations in \Soares.}. This will then
give rise to a massless representation of the massive superfields.

We will make progress towards the
second goal on a case by case basis, starting with the three point amplitude with
one massive state computed in \PSthreemass. Firstly, it will be simplified using BRST cohomology manipulations
in terms of the massive superfields. Subsequently, the massless SYM representation of the massive
superfields will be plugged in, allowing several further BRST cohomology identities for massless
expressions to be used. The end result expresses the three-point amplitude with one massive state in
terms of the massless four-point pure spinor superspace expression capturing the $\ap^2$
correction to the massless four-point open string tree amplitude. This result will be generalized
using the component expression of the massive partial tree amplitudes found in \massSweden\ to
linear combinations of $\ap^2$ tree amplitudes. Finally, the generalization of this relation will
be shown to follow from (or be compatible with) the factorization of the massless tree amplitudes
on its first massive residue. Our analysis of factorization is slightly unusual as the sum over
intermediate polarizations -- which usually require two amplitudes connected via a propagator -- will
be used in a single amplitude with the rule ${\underline k}\to i,j$ defined in \Hmap\ and \sumgb.
This allows us to directly relate massive and massless string amplitudes.

Throughout this paper, repeated vector indices are summed irrespective of their
downstairs/upstairs placement and we use the convention where the symmetrization or antisymmetrization over
$n$ indices
does not contain the normalization $1\over n!$.

\newnewsec\Vmasssec Vertex operators

Physical states in the pure spinor formalism at the mass level $n$
are defined as ghost number one vertex
operators in the cohomology of the pure spinor BRST charge
with
conformal weight $n$ at zero momentum \ICTP.

\newsubsec\Vmasslesssec Massless vertices

The unintegrated and integrated vertex operators describing the massless open string states
are given by \psf
\eqn\vertices{
V=\lambda^\alpha A_\alpha\,,\quad\quad U=\p\theta^\alpha A_\alpha+\Pi^mA_m
+2\alpha'd_\alpha W^\alpha+\alpha'N^{mn}F_{mn}\,.
}
The superfields $A_\alpha, A_m, W^\alpha$ and
$F_{mn}=\p_{[m}A_{n]}$ satisfy \refs{\wittentwistor,\treereview}
\eqn\SYMBRST{
\eqalign{
Q A_\beta +  D_\beta V &= (\g^m\l)_{\beta}A_m\,,\cr
QA_m &= (\l\g^m W) + \p_m V\,,
}\qquad
\eqalign{
Q W^\beta &= -\frac{1}{4}(\g^{mn}\l)^{\b}F_{mn},\cr
Q F_{mn} &=\p_m(\l\g_n W)- \p_n(\l\g_m W)\,,
}}
where $Q=\l^\a D_\a$ is the pure spinor BRST operator acting on $10$D superfields.
The length dimensions are chosen such that
\eqnn\massdim
$$\displaylines{
[\ap] = 2,\quad [V]=[U]=1,\quad
[A_\a] = \half\,,\quad
[A^m] = 0\,,\quad
[W^\a] = -\half\,,\quad
[F^{mn}] = -1\,\hfill\massdim\hfilneg\cr
[\l^\a]=[\t^\a] = \half,\quad [\p_m] = -1,\quad [d_\a] = -\half,\quad [\Pi^m] = 1,\quad [J]=0,
\quad[N^{mn}] =0\,.
}$$
By stripping off $\l^\a$ from \SYMBRST,
one obtains the equations of motion written in terms of the covariant derivative $D_\a$.
We will use below both forms of these equations interchangeably. For convenience, when referring
to a generic SYM
superfield labelled by $i$ we use the collective notation
\eqn\notSYM{
K_i \in \{A_\a^i,A_m^i, W^\a_i, F^{mn}_i\}\,.
}

\newsubsec\Vfirstmasssec Massive unintegrated vertex

The unintegrated vertex operator $V(z)$ containing the open-string massive states with $({\rm
mass})^2{=}1/\ap{=}-k^2$
was found in \BCpaper,
\eqn\VBC{
V(z)= [\l^\a[\p\t^\b B_{\a\b}]_0]_0 + [\l^\a[\Pi^m H^m_\a]_0]_0 + 2\ap [\l^\a[d_\b C^\b{}_\a]_0]_0
+ \ap[\l^\a[N^{mn} F_{\a mn}]_0]_0\,,
}
where the normal-ordering bracket $[AB]_0$ is defined in \NO. It was also shown in \BCpaper\ that
this vertex is BRST closed $QV=0$ when the superfields obey the equations of motion
\eqn\eomBC{
Q(\l B)_\a = (\l\g^m)_\a (\l H)_m\,,\quad
Q(\l H^m) = (\l\g^m C\l)\,,\quad
Q(C\l)^\a = {1\over4}(\l\g^{mn})^\a (\l F)_{mn}\,,
}
where we used the definitions \lfields\ and omitted the slightly
more complicated equation for $\l^\a F_{\a mn}$. The length dimension
of the massive superfields in \VBC\ is chosen to be
\eqn\massdimV{
[V]=2,\quad
[B_{\a\b}] = 1,\quad
[H_{m\a}] = \half\,,\quad
[C^\b{}_\a] = 0\,,\quad
[F_{\a mn}] = -{1\over2}\,.
}

\newsubsec\VfirstmassOPEsec Massive vertex from the OPE of massless vertices

Massive vertex operators appear in the regular terms of OPEs of massless vertices  \FMS.
This will allow us to construct the first-level massive unintegrated vertex operator in terms of the massless superfields
\refs{\Chakrabarti, \Koh} as
\eqn\massV{
V(z) =  \oint_z dw U_1(w) V_2(z),\qquad 2\ap (k_1\cdot k_2) = -1\,,
}
where the condition
\eqn\firstcond{
2\ap (k_1\cdot k_2) = -1\,,
}
ensures the correct conformal weight one for the vertex $V(z)$.

\paragraph{OPE of massless vertices} It is easy to see from the OPE expansion \opeAB\ that
\eqn\countV{
V(w)=\oint dz U_1(z)V_2(w)=[U_1V_2]_1,
}
where the bracketed notation for the OPEs and normal ordering is reviewed in the appendix~\opeapp.
Using the normal ordered massless vertex operators\foot{Note that there is no normal ordering
ambiguity in the massless vertices due to the SYM equations of motion in the Lorenz gauge.
Nevertheless, we write the normal ordering brackets in order to use the OPE formulas from the
vertex operator algebra axioms of the appendix~\opeapp.}
\eqnn\masslessU
\eqnn\masslessV
$$\eqalignno{
U_1(z) &=
[\p\t^\a A_\a^1]_0(z)
+ [\Pi^m A^1_m]_0(z)
+ 2\ap[d_\a W^\a_1]_0(z)
+ \ap [N^{mn}F_1^{mn}]_0(z)\,, &\masslessU\cr
V_2(w) &= [\l^\a A_\a^2]_0(w)\,,&\masslessV
}$$
we get (we write $[AB]_0=[AB]$ when convenient to avoid cluttering)
\eqnn\UoVt
$$\eqalignno{
V(w) &=[\l^\b[\p\t^\a (A^1_\a A^2_\b)]](w)
+[\l^\b[\Pi^m (A^1_m A^2_\b)]](w) - 2\ap i k_2^n[\l^\b (\p A^1_n A^2_\b)](w) \cr
&+[\l^\b[d_\a (W_1^\a A^2_\b)]](w) - [\l^\b (\p W_1^\a D_\a A^2_\b)](w)\cr
&+[N^{mn}[\l^\b(F_1^{mn}A_\b^2)]](w)
+\half(\g^{mn})^\b{}_\d[\l^\d (\p F_1^{mn}A_\b^2)](w) &\UoVt\cr
}$$
The appendix \detailsapp\ contains more details of this calculation.

Note that the factors of $(K_1 K_2)(w)$ on the right-hand side are considered a single
operator.
For example, the term $[d_\a (W_1^\a A_\b^2)]_0(w)$
is of the form $[AB]_0(w)$ with
$A=d_\a(w)$ and $B=(W_1^\a A^2_\b)(w)=W_1^\a(\t)e^{ik_1\cdot X(w)}A^2_\b(\t)e^{ik_2\cdot
X(w)}=W_1^\a(\t)A^2_\b(\t)e^{ik\cdot X(w)}$, with $k=k_1+k_2$.

\newsubsec\MassOPEsec Massive superfields in the OPE gauge

After expanding $\p K_1 = \p\t^\a D_\a K_1 + \Pi^m ik_1^m K_1$ to rewrite factors like
$(\p A_1^n A^2_\b)$ as $[\Pi^m ik^m_1 (A^n_1 A^2_\b)]_0 + [\p\t^\a (D_\a A^n_1 A^2_\b)]_0$
and
using \firstcond\ together with the SYM equations of
motion we get (omitting the worldsheet position $w$ from the right-hand side)
\eqn\UoVtvertex{
V(w)= [\l^\a[\p\t^\b B_{\a\b}]] + [\l^\a[\Pi^m H^m_\a]] + 2\ap [\l^\a[d_\b C^\b{}_\a]]
+ \ap[N^{mn}[\l^\a F_{\a mn}]]
}
where the massive superfields can be read off to be
\eqnn\Balbe
\eqnn\Hmal
\eqnn\Cab
\eqnn\Famn
$$\eqalignno{
B_{\a\b}&= -2\ap i k_2^m(\g^m W_1)_\b A^2_\a - \ap i k_1^m (\g^n W_1)_\b
(\g^{mn}A_2)_\a - {\ap\over2}F_1^{mn}(\g^{mn}D)_\b A^2_\a\,, \qquad &\Balbe\cr
H^m_\a &=
A^{1}_m A^{2}_\a
	+ 2\ap k^1_m(k^2\cdot A^{1}) A^{2}_\a
	-2 \ap  ik^1_m W^\b_1D_\b A^{2}_\a
	- {\ap\over2}ik^1_m F^{1}_{np}(\gamma^{np}A_2)_\a\,, &\Hmal\cr
C^\b{}_\a &=W_1^\b A^{2}_\a\,, &\Cab\cr
F_{\a mn} &=F^{1}_{mn}A^{2}_\a\,. &\Famn
}$$
For reasons to become clear in section~\bcgaugesec,
this representation of the massive superfields in terms of massless SYM superfields will be called
the {\it OPE gauge}. Their length dimensions are easily found to agree with \massdimV.

\paragraph{Massive equations of motion}
Defining the contraction of the massive superfields with a pure spinor
\eqn\lfields{
\l^\a B_{\a\b}=(\l B)_\b,\quad
\l^\a H^m_\a = (\l H^m)\,,\quad
C^\b{}_\a \l^\a = (C\l)^\b\,,\quad
\l^\a F_{\a mn} = (\l F)_{mn}\,,
}
and using the linearized SYM equations of motion \SYMBRST\ one
readily finds the equations of motion of the massive superfields in terms of the BRST charge
$Q=\l^\a D_\a$,
\eqnn\Beom
\eqnn\Heom
\eqnn\Ceom
\eqnn\Feom
$$\eqalignno{
Q(\l B)_\a &= (\l\g^m)_\a (\l H)_m &\Beom\cr
Q(\l H^m) &= (\l\g^m C\l)&\Heom\cr
Q(C\l)^\a &= {1\over4}(\l\g^{mn})^\a (\l F)_{mn}&\Ceom\cr
Q(\l F)_{mn} &= ik^m_1(\l\g^n W_1)V_2 - ik^n_1(\l\g^m W_1)V_2 &\Feom
}$$
More details can be found in the appendix~\eomapp.

\newsubsubsec\brstv BRST invariance of massive vertex

Recall that the BRST charge is
\eqn\Qcharge{
Q=\oint dz j(z)
}
where $j(z)=\l^\a(z) d_\a(z) = [\l^\a d_\a]_0(z)$ is the BRST current.
We will evaluate the BRST variation of the massive unintegrated vertex in two different ways:
directly from the definition \countV\ and using its explicit realization
\UoVtvertex.

\paragraph{BRST variation from the definition}
The massless vertices satisfy \refs{\psf,\treereview}
\eqnn\jmassless
$$\eqalignno{
QU_1(w) &= \oint dz j(z) U_1(w) = [jU_1]_1(w) = \p V_1(w)\,,&\jmassless\cr
QV_2(w) &= \oint dz j(z) V_2(w) = [jV_2]_1(w) = 0\,.
}$$
Therefore, the BRST variation of the first massive unintegrated vertex operator
\countV\ yields
\eqnn\QVmass
$$\eqalignno{
QV(w) 
&= [jV]_1(w) =[j[U_1V_2]_1]_1 &\QVmass\cr
&= [U_1[jV_2]_1]_1 + [[jU_1]_1 V_2]_1 = [\p V_1 V_2]_1 = 0,
}$$
where the second line follows from \deriv\ and we used \pAB\ in the last equality.
So the massive unintegrated vertex \countV\ is BRST closed.


\paragraph{Evaluation using superfields}
The explicit computation of $[jV]_1$ with $V$ given by \UoVtvertex\
is a bit tedious but
straightforward. Using the identities \ABC\ and \ABCt\ gives
\eqnn\brstvar
$$\eqalignno{
QV=[jV]_1 &=[\p\l^\a[\l^\b B_{\b\a}]]
- [\p\t^\b[\l^\g[\l^\a D_\a B_{\g\b}]]]
+ [[\l^\a\p\t^\g][\l^\b H^m_\b]]\g^m_{\a\g}\cr
& - [[\l^\a \Pi^m][\l^\b C^\g{}_\b]]\g^m_{\a\g}
+[\Pi^m[\l^\b[\l^\a D_\a H^m_\b]]]\cr
&-{\ap\over2}[\l^\b[[d_\a\l^\g]F_{\b mn}]](\g^{mn})^\a{}_\g
-2\ap[d_\g[\l^\b[\l^\a D_\a C^\g{}_\b]]]\cr
&+\ap[N^{mn}[\l^\a[\l^\b D_\b F_{\a mn}]]]\,. &\brstvar
}$$
In order to compare terms we need to rewrite all nested brackets in a canonical order, say from
right to left as in $[A[B[CD]]]$. After some work using the
identities \ABn\ to \ApB\ one obtains
\eqnn\canonic
$$\eqalignno{
[\l^\b[[d_\a\l^\g]F_{\b mn}]] &=
[d_\a[\l^\b[\l^\g F_{\b mn}]]] + [\p\l^\g[\l^\b(D_\a F_{\b mn})]]\,, &\canonic\cr
[[\l^\a\p\t^\g][\l^\b H^m_\b]] &= [\p\t^\g[\l^\a [\l^\b H^m_\b]]]\,,\cr
[[\l^\a \Pi^m][\l^\b C^\g{}_\b]] &= [\Pi^m[\l^\a[\l^\b C^\g{}_\b]]]
-2\ap [\p\l^\a[\l^\b \p^m C^\g{}_\b]]\,.\cr
}$$
Plugging \canonic\ into \brstvar\ and noticing that there are no normal
ordering ambiguities among
$\l^\a$, $\p\t^\a$ and the massive superfields
leads to
\eqnn\QVvertex
$$\eqalignno{
QV &=[\p\l^\a\l^\b S^1_{\a\b}]
- [\p\t^\a[\l^\b\l^\g S^2_{\a\b\g}]]
+ [\Pi^m[\l^\a\l^\b S^3_{m\a\b}]] &\QVvertex\cr
&+2\ap [d_\a[\l^\b\l^\g S^{4\a}_{\b\g}]]
+ \ap [N^{mn}[\l^\a\l^\b S^5_{\a\b mn}]]\,,
}$$
where
\eqnn\QVu
\eqnn\QVd
\eqnn\QVt
\eqnn\QVq
\eqnn\QVc
$$\eqalignno{
\p\l^\a\l^\b S^1_{\a\b}&= (\l B\p\l) +2\ap (\p\l\g^m \p^m C\l) +{\ap\over2}(\p\l\g^{mn}D)
(\l F)_{mn} &\QVu\cr
&=-\ap ik^m_1(\l\g^n W_1)(\p\l\g^{mn}A_2)\,,\cr
\l^\b\l^\g S^2_{\a\b\g}&= Q (\l B)_{\a} -(\l H^m)(\l\g^m)_{\a} =0\,,&\QVd\cr
\l^\a\l^\b S^3_{m\a\b}&=Q (\l H^m) - (\l\g^m C\l) =0\,,&\QVt\cr
\l^\b\l^\g S^{4\a}_{\b\g}&=Q (C\l)^\a - {1\over4}(\l\g^{mn})^\a (\l F)_{mn} =0\,,&\QVq\cr
\l^\a\l^\b S^5_{\a\b mn}&= Q (\l F)_{mn} =
ik^m_1(\l\g^n W_1)V_2 - ik^n_1(\l\g^m W_1)V_2\,. &\QVc\cr
}$$
The simplification in \QVu\ follows from the linearized SYM
equations of motion and the Dirac equation after plugging in the expression \Balbe.
The vanishing of the middle three lines follows from the BRST equations of
motion \Beom, \Heom\ and \Ceom.
Therefore,
\eqn\QVtmp{
QV=-\ap ik^m_1(\l\g^n W_1)(\p\l\g^{mn}A_2) - 2\ap i[N^{mn}[\l^\a\l^\b]]\g^m_{\b\g}k^n_1 W_1^\g A_\a^2\,,
}
where we pulled the superfields out of the normal ordering bracket as they do not have worldsheet
singularities with the operators in $[N^{mn}[\l^\a\l^\b]]$. The identity from \BCpaper
\eqn\Nlalaid{
[N^{mn}[\l^\a\l^\b]]\g^m_{\b\g} = \half[J[\l^\a\l^\b]]\g^n_{\b\g}
+{5\over2}\l^\a(\g^n\p\l)_\g
+ \half(\l\g^{mn})^\a(\g^m\p\l)_\g\,,
}
rederived in \tmpNlala, implies that
\eqn\QVtmpI{
QV=-\ap ik^m_1(\l\g^n W_1)(\p\l\g^{mn}A_2) - \ap i k^n_1 (\p\l\g^m W_1) (\l\g^{mn} A_2) = 0\,,
}
where the Dirac equation eliminates the first two terms on the right-hand side of \Nlalaid\ when
plugged into \QVtmp\ and we
used $(\l\g^m)_\a(\p\l\g^m)_\b + (\l\g^m)_\b(\p\l\g^m)_\a = 0$.
Therefore the unintegrated massive vertex \UoVtvertex\ is BRST closed, $QV=0$.

\newsubsec\bcgaugesec Massive superfields in the Berkovits-Chandia gauge

We have seen above that the OPE calculation leads to an unintegrated massive vertex operator of the form
\eqn\Vvertex{
V(w)= [\l^\a[\p\t^\b B_{\a\b}]] + [\l^\a[\Pi^m H^m_\a]] + 2\ap [\l^\a[d_\b C^\b{}_\a]]
+ \ap[N^{mn}[\l^\a F_{\a mn}]]\,,
}
with coefficients given in \Balbe\ to \Famn.
It does not contain the fields $\p\l^\a$ and $J$
that would otherwise be present in the most general form of $V$ of conformal weight one and ghost-number one.
This parameterization in \Balbe\ to \Famn\ was called the {\it OPE gauge}.

\paragraph{Berkovits-Chandia gauge}
As shown in \BCpaper, the gauge invariance $\d V = Q\Omega$ can be exploited to obtain a new
parameterization for the superfields such that
\eqnn\BCgauge
$$\displaylines{
B_{\a\b} = \g^{mnp}_{\a\b}B_{mnp}\,,\qquad \p^m B_{mnp} = 0\,,\qquad
\g^{m\a\b}H_{m\b} = 0\,,\qquad \p^m H_{m\a} = 0,\hfil\BCgauge\hfilneg\cr
C^\a{}_\b =
{1\over4}(\g^{mpnq})^\a{}_\b\p_{m}B_{npq}\,,\qquad
\g^{m\a\b}F_{\a mn} = 0\,,
}$$
which we will call the {\it Berkovits-Chandia gauge}. As a side note,
the normal ordering identity \frombais\ yields
$[N^{mn}[\l^\a F_{\a mn}]_0]_0=[\l^\a[N^{mn}F_{\a mn}]_0]_0 +\half[(\g^{mn}\p\l)^\a F_{\a mn}]_0$
and therefore constraint $\g^{m\a\b}F_{\a mn} = 0$ implies that
\eqn\NlaABC{
[N^{mn}[\l^\a F_{\a mn}]_0]_0
= [\l^\a[N^{mn}F_{\a mn}]_0]_0\,,
}
a relation that will be exploited later in \Vmass.

This same gauge fixing will now be done starting from the vertex in the OPE gauge.

\paragraph{Gauge-fixed massive superfields}
The gauge invariance of the massive vertex $\d V=Q\Omega$ with the most general
superfield $\Omega$
of conformal weight one and ghost number zero,
\eqn\Omegadef{
\Omega = [\partial\theta^\alpha\Omega_{1\alpha}]_0
	+ [d_\alpha\Omega^\alpha_2]_0
	+ [\Pi^m\Omega_{3m}]_0
	+ [J\Omega_4]_0
	+ [N^{mn}\Omega_{5mn}]_0\,,
}
will now be exploited to go from massive superfields in the OPE gauge \Balbe-\Famn\ to massive
superfields in the Berkovits-Chandia gauge satisfying \BCgauge.

The BRST variation $Q\Omega = [j\Omega]_1$, where $j=[\l^\a d_\a]_0$ is the BRST current, reads
\eqnn\Qom
$$\eqalignno{
Q\Omega &=
	[\p\theta^\b\lambda^\a]_0 \Bigl(- D_\a\Omega_{1\b} 
    +\g^m_{\a\b}\Omega_{3m}\Bigl)
	+[\Pi^m\lambda^\a]_0 \Bigl( D_\a\Omega_{3m}
    - \frac{1}{2\ap }\g^m_{\a\b}\Omega^\b_2\Bigl) &\Qom\cr
    &+ [d_\b\lambda^\a]_0 \Bigl( -D_\a\Omega^\b_2
    - \frac{1}{2}(\g^{mn})^\b_{\ \a}\Omega_{5mn}
    - \delta^\b_\a\Omega_4\Bigl) + [J\lambda^\a]_0 D_\a\Omega_4 \cr
 &+ [N^{mn}\lambda^\a]_0 D_\a\Omega_{5mn}+ \p\lambda^\a \Bigl(\Omega_{1\a}
    + \g^m_{\a\b}\p_m\Omega^\b_2
    - D_\a\Omega_4
    - \frac{1}{2}(\g^{mn})^\b_{\ \a}D_\b\Omega_{5mn}\Bigl)\,.
}$$
However, the
gauge variations of the massive superfields following from \Qom\ need to be modified by a vector-spinor
parameter $\Lambda^\b_n$ to account for the
constraint \BCpaper
\eqn\Nlat{
	[N^{mn}\lambda^\a]_0\g_{m\a\b}
	-\frac{1}{2}[J\lambda^\a]_0\g^n_{\a\b}
	= 2\p\lambda^\a\g^n_{\a\b}\,.
}
The resulting gauge transformations are given by
\eqnn\ghty
$$\eqalignno{
    \delta B_{\a\b} & =
    - D_\a\Omega_{1\b}
    +\g^m_{\a\b}\Omega_{3m}\,, &\ghty\cr
    \delta H_{m\a} & =
     D_\a\Omega_{3m}
    - \frac{1}{2\ap }\g^m_{\a\b}\Omega^\b_2\,,\cr
    \delta C^\b_{\ \a} &=
    -\frac{1}{2\ap }D_\a\Omega^\b_2
    - \frac{1}{4\ap }(\g^{mn})^\b_{\ \a}\Omega_{5mn}
    - \frac{1}{2\ap }\delta^\b_\a\Omega_4\,,\cr
    \delta F_{\a mn} & =
     \frac{1}{\ap }D_\a\Omega_{5mn}
    + \g_{m\a\b}\Lambda^\b_n
    - \g_{n\a\b}\Lambda^\b_m\,,
}$$
and it is easy to determine the length dimensions of the gauge parameters
\eqn\lengthOmega{
[\Omega^1_\a]={3\over2},\quad [\Omega^2_\a]={5\over2}, \quad[\Omega_3^m] = 1, \quad[\Omega_4] =
2,\quad [\Omega_5^{mn}] = 2,\quad[\Lambda^\a_m]=-\half.
}
Since there are no massive superfields proportional to $J$ and $\p\l^\a$ in the massive vertex
$V$ in the OPE gauge \Vvertex,
the following constraints need to be satisfied as well
\eqnn\addco
$$\eqalignno{
    0 &=
     D_\a\Omega_4
    - \ap \g^m_{\a\b}\Lambda^\b_m &\addco\cr
    0 & =      \Omega_{1\a}
    + \g^m_{\a\b}\p_m\Omega^\b_2
    - D_\a\Omega_4
    - \frac{1}{2}(\g^{mn})^\b_{\ \a}D_\b\Omega_{5mn}
    -4\ap \g^m_{\a\b}\Lambda^\b_m\,.
}$$
Note that we can eliminate the term involving $\Lambda^\a_m$ from
the above two equations to arrive at a single condition
\eqn\anotherconstraint{
	\Omega_{1\a}
	+ \g^m_{\a\b}\p_m\Omega^\b_2
	- 5 D_\a\Omega_4
	- \frac{1}{2}(\g^{mn})^\b_{\ \a}D_\b\Omega_{5mn} = 0\,.
}

\newsubsubsec\Balbesec $B_{\a\b}$ in the Berkovits-Chandia gauge

After the gauge transformation, the superfield $B_{\a\b}$ takes the form 
\eqn\Balbetrans{
B'_{\a\b} \;= \;B_{\a\b}  - D_\a\Omega_{1\b} 
    +\g^m_{\a\b}\Omega_{3m}
}
Now, the bispinor $B'_{\a\b}$ decomposes into a one-, three- and five-form
parts
\eqnn\bprime
$$\eqalignno{
	B'_{\a\b}&= \frac{1}{16}\g^{m}_{\a\b}\g_{m}^{\sigma\tau}
	\Bigl(
	B_{\sigma\tau}-D_\sigma\Omega_{1\tau}
	\Bigl)
	\;+\;\g^m_{\a\b}\Omega_{3m} &\bprime\cr
	& +\;\frac{1}{96}\g^{mnp}_{\a\b}\g_{mnp}^{\sigma\tau}
	\Big(
	B_{\sigma\tau}-D_\sigma\Omega_{1\tau}
	\Big)\cr
	&\ +\;\frac{1}{3840}\g^{mnpqr}_{\a\b}\g_{mnpqr}^{\sigma\tau}
	\Big(
	B_{\sigma\tau}-	D_\sigma\Omega_{1\tau}
	\Big)
}$$
However, the five-form part is BRST exact
$\lambda^\a\lambda^\b\Big(B_{\a\b}-D_\a\Lambda_\b\Big) = 0$
as shown in \exproof.
So if we choose the
gauge parameter $\Omega_{1\tau} = \Lambda_\tau$, where $\Lambda_\tau$ is defined in \Lamdef,
we can eliminate the 5-form piece from
\bprime. Further, choosing
\eqn\omthreeA{
	\Omega_{3m} = -\frac{1}{16}\g_m^{\sigma\tau}
	\Big(
	B_{\sigma\tau} - D_\sigma\Lambda_\tau
	\Big)
}
eliminates the $1$-form
and $B'_{\a\b}$ becomes
\eqn\Bprimetmp{
	B'_{\a\b} =
	\frac{1}{96}\g^{mnp}_{\a\b}\g_{mnp}^{\sigma\tau}
	\Big(
	B_{\sigma\tau}-D_\sigma\Lambda_{\tau}
	\Big)
}
Simplifying the above expression using \Lamdef, we arrive at the result
\eqn\Bprimetmpd{
	B'_{\a\b} = \g^{mnp}_{\a\b}B'_{mnp}
}
where,
\eqnn\bpmnp
$$\eqalignno{
	96B'_{mnp} &=
	 \Bigl[(A_1\g_{mnp}A_2)+8\ap A^{1}_{[m}F^{2}_{np]}
	-4\ap (W_1\g^{mnp}W_2)\cr
	&\ + 2i\ap k^{1}_{[m}(A_1\g_{np]}W_2)
	+ 4i\ap  k^1_q(A_1\g_{mnpq}W_2) + (1\leftrightarrow2)\Bigr]&\bpmnp
}$$
Note that the above expression of $B'_{mnp}$ is invariant under the exchange of massless-particle
labels. However, this is still not in the Berkovits-Chandia gauge since it does not satisfy the condition
$k^m B'_{mnp}=0$, where $k^m=k^m_1+k^m_2$. To satisfy this condition, we note that we are still allowed to change
$\Omega_{1\a} = \Lambda_\a$ by shifting with any $\Phi_\a$ which satisfies
$Q(\lambda^\a\Phi_\a) = 0$.
One can show that the following expressions are
BRST closed (and also BRST exact)
\eqnn\alsoexact
$$\eqalignno{
	\lambda^\a\Phi_{1\a} &=
	ik^2_mA^{1}_mV_2
	+ ik^1_mV_1 A^{2}_m
	+ A^{1}_m(\l\g^mW_{2})
	+ (\lambda\g^m W_{1}) A^{2}_m = Q(A_1\cdot A_2) &\alsoexact\cr
	\lambda^\a\Phi_{2\a} &=
	ik^1_n(\lambda\g^mW_{1}) F^{2}_{mn}
	+ ik^2_nF^{1}_{mn}(\l\g^mW_{2}) = -\half Q(F_1^{mn}F_2^{mn})\,.
}$$
Since the above expressions are BRST closed, we can modify the gauge parameters $\Omega_{1\a}$
and $\Omega_{3m}$ by terms involving $\Phi_{1\a}$ and $\Phi_{2\a}$. Choosing
\eqnn\omthree
$$\eqalignno{
	\Omega_{1\a} &= \Lambda_\a + 2\ap \Phi_{1\a} + \frac{4}{3}\ap ^2\Phi_{2\a}\,,&\omthree\cr
	\Omega_{3m} &= -\frac{1}{16}\g_m^{\sigma\tau}
	\Big(B_{\sigma\tau} - D_\sigma\bigl(\Lambda_\tau+2\ap \Phi_{1\tau}+\frac{4}{3}\ap ^2\Phi_{2\tau}\bigl)
	\Big)\,,
}$$
we arrive at the following expression of $B'_{mnp}$
\eqnn\bpabc
$$\eqalignno{
B'_{mnp} &= \frac{1}{18}\ap (W_1\g_{mnp}W_2)
+\frac{1}{9}\ap ^2k^{1}_{[m}k^{2}_{n}(W_{1}\g_{p]}W_{2})
+\frac{1}{18}i\ap^2\Big[k^{2q}F^{1}_{q[m}F^{2}_{np]} +(1\leftrightarrow 2)
\Big]\cr
&={1\over9}\ap^2 (W_1\g^{k^1 k^2 mnp}W_2)
+\frac{1}{18}\ap^2\Big[ik^{2q}F^{1}_{q[m}F^{2}_{np]} +(1\leftrightarrow 2) \Big]\,, &\bpabc
}$$
where we used the shorthand $(W_1\g^{k^1 k^2 mnp}W_2)= k^1_a k^2_b (W_1\g^{abmnp}W_2)$ in the
second line.
The equivalence between the first and second lines of \bpabc\
follows from the Dirac equation and $2\ap k_1\cdot k_2=-1$. It is straightforward to show that
\bpabc\ indeed satisfies $k^{m}B'_{mnp} = 0$.

The explicit form of the gauge parameters defined
in \omthree\ in terms of the massless superfields is given by
\eqnn\OmOneOmThree
$$\eqalignno{
\Omega_{1\a}&=\half\ap F^{mn}_1(\g_{mn}A_2)_\a
+ 2\ap (\g_m W_1)_\a A^m_2 &\OmOneOmThree\cr
& + {4\over3}i\ap^2 k_n^1(\g_m W_1)_\a F_2^{mn}
+ {4\over3}i\ap^2 k_n^2(\g_m W_2)_\a F_1^{mn}\,,\cr
	\Omega_{3m} &= 2i\ap k^1_m(W_1A_2)
	+ \ap (W_{1}\g_m W_{2})
	+ 2\ap F^{1}_{mn}A^{n}_2\cr
	& + \frac{2}{3}i\ap ^2k^1_pF^{1}_{mn}F^{np}_2
	+ \frac{2}{3}i\ap ^2k^2_pF^{2}_{mn}F^{np}_1\,.
}$$

\newsubsubsec\Hmsec $H_{m\a}$ in the Berkovits-Chandia gauge

We next consider the superfield $H_{m\a}$. After the gauge transformation, it becomes
\eqn\HpdefA{
H'_{m\a} = H_{m\a} +D_\a\Omega_{3m}
	-\frac{1}{2\ap }\g^m_{\a\b}\Omega^\b_2\,.
}
It is easy to see that the condition $\g^{m\a\b}H'_{m\a}=0$ is satisfied provided we choose
\eqnn\omtwo
$$\eqalignno{
	\Omega^\a_2 &=
	\frac{\ap }{5}\g^{m\a\b}
	\Big(
		H_{m\b}
		+ D_\b\Omega_{3m}
	\Big)\cr
&=
	-4i\ap ^2k^1_mW^{\a}_1 A^{m}_2
	-\frac{4}{3}i\ap ^2k^1_n A^{1}_m (\g^{mn}W_2)^\a
	+\frac{2}{3}i\ap ^2k^2_n(\g^{mn}W_1)^\a A^{2}_m\,,&\omtwo
}$$
which implies ($k^m_{12}=k^m_1+k^m_2$)
\eqnn\Hpdef
$$\eqalignno{
	H^{'m}_{\a} &=
	 {i\ap\over6}\Bigl(-5i  F_{1}^{mn}(\g_nW_{2})_\a
	-2 k_{12}^mA^{1}_n(\g^n W_2)_\a
	+  k^1_pA^{1}_n(\g^{mnp}W_{2})_\a
	\cr
	&\ \quad
	-4\ap k_{12}^m(k^2\cdot A^{1})k^1_n(\g^nW_{2})_\a \
	+\ (1\leftrightarrow 2) \Bigr)\,. &\Hpdef\cr
}$$
It is straightforward to prove that \Hpdef\ satisfies the transversality condition
$k^mH'_{m\a} = 0$ where $k^m = k^m_1+k^m_2$. In addition, a long calculation using the massless
equations of motion \SYMBRST\ (stripping off the pure spinor) and the constraint $2\ap k_1\cdot k_2
= -1$ reveals that
\eqn\DBmnp{
H'_{m\a} = {3\over7}(\g^{np})_\a{}^\b D_\b B'_{mnp}\,.
}

\newsubsubsec\Cabfix $C^\b{}_\a$ in the Berkovits-Chandia gauge

Next, we consider the superfield $C^\b_{\;\;\a}$. Its gauge transformation implies
\eqn\cab{
C'^\b_{\;\;\a}=C^\b_{\;\;\a}-\frac{1}{2\ap }D_\a\Omega^\b_2
    - \frac{1}{4\ap }(\g^{mn})^\b_{\ \a}\Omega_{5mn}
    - \frac{1}{2\ap }\delta^\b_\a\Omega_4\,.
}
After applying the Fierz identity with respect to the indices $\a$ and $\b$
one can eliminate the zero- and two-form parts by choosing
\eqnn\elimzt
$$\eqalignno{
\Omega_4 &= \frac{\ap }{8}
	\Big(
		C^\a_{\ \a}
		- \frac{1}{2\ap }D_\a\Omega^\a_2
	\Big)\,,&\elimzt\cr
 \Omega_{5mn} &=  \frac{\ap }{8}(\g_{mn})^\a_{\ \b}
	\Big(
		-C^\b_{\ \a}
		+\frac{1}{2\ap }D_\a \Omega^\b_2
	\Big)\,.
}$$
In terms of the massless superfields, their explicit expressions are given by
\eqnn\omone
$$\eqalignno{
	\Omega_4 &=-{\ap^2\over6}F^{mn}_1F_{mn}^2\,,&\omone\cr
	\Omega_{5mn} &=
	\ap^2\Bigl( F^1_{a[m}F^2_{n]a} -2 F^1_{mn}(ik^1\cdot A^{2})
	+\frac{3}{2}ik^1_{[m}(W_{1}\g_{n]}W_{2})
	+\frac{1}{2}ik^2_{[m}(W_{1}\g_{n]}W_{2})\Bigr)\,,
%
}$$
with $2\ap k_1\cdot k_2 = -1$. Plugging these into \cab, we find
($k_{12}^m = k_1^m + k_2^m$),
\eqn\Cpab{
	C'^\b{}_\a = {\ap\over6}(\g^{mnpq})^\b_{\ \a}
	\Bigg(
		\frac{1}{12}i k^{12}_m(W_{1}\g_{npq}W_{2})
		+ k^1_mk^2_nA^{1}_pA^{2}_q
	\Bigg)\,.
}
It is easy to see that the above expression is equivalent to ($k^m=k^m_{12}$)
\eqn\reqCp{
	C'^\b{}_\a = \frac{1}{4}ik_m(\g^{mnpq})^\b_{\ \a}B'_{npq}\,,
}
with $B'_{mnp}$ given in equation \bpabc.

As a consistency check, we find that the gauge superfield parameters
$\Omega_{1\a}$, $\Omega^\a_2$, $\Omega_4$ and $\Omega_{5mn}$ fixed in \omthree,
\omtwo\ and \omone\ satisfy the constraint \anotherconstraint.

\newsubsubsec\Famnfixsec $F_{\a mn}$ in the Berkovits-Chandia gauge

Finally, we consider the superfield $F_{\a mn}$. After the gauge transformation, it takes the form
\eqn\forone{
F'_{\a mn}=F_{\a mn} +\frac{1}{\ap }D_\a\Omega_{5mn}
    + \g_{m\a\b}\Lambda^\b_n
    - \g_{n\a\b}\Lambda^\b_m\,.
}
Requiring the constraint $\g^{\a\b}_m F'_{\b mn}=0$ and using
\addco\ and $\{\g^m,\g^n\}=2\eta^{mn}$
implies that
\eqn\Laln{
\Lambda_n^\a = -{1\over 8\ap}\Bigl( \ap \g_m^{\a\b}F_{\b mn} + (\g^m D)^\a\Omega^5_{mn}+ (\g^n
D)^\a\Omega_4\Bigr)\,.
}
Using the expressions of $F_{\a mn}$ from \Famn\ and $\Omega_4$ and $\Omega_{5 mn}$ from \omone,
we get
\eqnn\lamsol
$$\eqalignno{
	\Lambda^\a_m &=	- \frac{43}{96}A^{1}_mW^{\a}_2
	+ \frac{53}{96}W^\a_1 A^{2}_m -\frac{1}{96}A^{1}_n(\g^{mn}W_{2})^\a
	-\frac{1}{96}(\g^{mn}W_{1})^\a A^{2}_n &\lamsol\cr
	&
	-\frac{43}{48}\ap k^1_m(k^2\cdot A^{1})W^{\a}_2
	+ 2\ap k^1_mW^{\a}_1(k^1\cdot A^{2})
	+\frac{53}{48}\ap k^2_mW^\a_1(k^1\cdot A^{2})\cr
	&
	-\frac{13}{16}\ap k^1_mk^1_nA^{1}_p(\g^{np}W_{2})^\a
	+\frac{3}{16}\ap k^2_m(\g^{np}W_{1})^\a k^2_nA^{2}_p-\frac{3}{8}\ap
	k^2_mk^1_nA^{1}_p(\g^{np}W_{2})^\a\cr
	&
	+\frac{5}{8}\ap k^1_m(\g^{np}W_{1})^\a k^2_nA^{2}_p
	-\frac{1}{48}\ap (k^2\cdot A^{1})k^1_n(\g^{mn}W_{2})^\a
	-\frac{1}{48}\ap k^2_n(\g^{mn}W_{1})^\a(k^1\cdot A^{2})\,,
}$$
which indeed satisfies the first equation in \addco.
Finally, we find $F'_{\a mn}$ in the Berkovits-Chandia gauge as
\eqnn\Fpamn
$$\eqalignno{
	F'_{\a mn} &=
	- \frac{1}{24}A^{1}_{[m}(\g_{_{n]}}W_{2})_\a
	- \frac{1}{48}A^{1}_p(\g^{mnp}W_{2})_\a
	+ \frac{3}{8}\ap k^1_{[m}k^2_{n]}A^{1}_p(\g^pW_{2})_\a &\Fpamn\cr
	&
	+ \frac{7}{16}\ap k^1_{[m}A^{1}_{n]}k^1_p(\g^pW_{2})_\a
	- \frac{1}{12}\ap (k^2\cdot A^{1})k^1_{[m}(\g_{n]}W_{2})_\a\cr
	&
	+ \frac{3}{8}\ap k^2_{[m}A^{1}_{n]}k^1_p(\g^pW_{2})_\a
	+ \frac{1}{16}\ap k^1_pA^{1}_q k^1_{[m}(\g_{n]pq}W_{2})_\a\cr
	&
	+ \frac{1}{8}\ap k^1_pA^{1}_qk^2_{[m}(\g_{n]pq}W_{2})_\a
	- \frac{1}{24}\ap (k^2\cdot A^{1})k^1_p(\g^{mnp}W_{2})_\a\cr
	&
	+ (1\leftrightarrow 2)\,,
}$$
which satisfies ($[mn] = mn-nm$, $k^m=k_1^m+k_2^m$)
\eqn\hdefone{
	F'_{\a mn} = \frac{1}{16}
	\Big(
		7ik_{[m}H'_{n]\a}
		+ ik_q(\g_{q[m})_\a^{\ \b}H'_{n]\b}
	\Big)\,.
}
Furthermore, using the equation of motion for $H^{\prime m}_\a$ \masstheta,
\eqn\DH{
D_\a H^{\prime m}_\b = -{9\over2}G'_{mn}\g^n_{\a\b}
- {3\over2}\p_a B'_{bcm}\g^{abc}_{\a\b}
+ {1\over4}\p_a B'_{bcd}\g^{mabcd}_{\a\b}\,,
}
where $G'_{mn}{=}-{1\over144}\big( (D\g^m H'^n)+(D\g^n H'^m)\big)$,
several gamma matrix
identities and
\eqnn\interm
$$\eqalignno{
\p_p C'^\a{}_\d\g_p^{\d\b} &= -{1\over4\ap}(\g^{mnp})^{\a\b}B'_{mnp}\,,&\interm\cr
(\l\g^{mnabc}\l)B'_{abc} &= 2\ap\bigl(
\p_p(\l\g^m C'\g^n\g^p\l)
- \p_p(\l\g^n C'\g^m\g^p\l)\bigr)\,,
}$$
we obtain
\eqnn\QlaFmnEom
$$\eqalignno{
Q(\l F')_{mn} &= {1\over2}\p_{[m}(\l\g_{n]} C'\l)
-{1\over16}\p^p(\l\g_{[m} C'\g_{n]p}\l)\,, &\QlaFmnEom\cr
%
}$$
where  $(\l\g^m C'\g^{np}\l) = (\l\g^m)_\a C'^\a{}_\b (\g^{np}\l)^\b$.

\newsubsubsec\sumsec Massive superfields in the BC gauge: summary

In summary, the massive superfields in the Berkovits-Chandia gauge
written in terms of massless superfields\foot{These results were firstly announced in November 2023 \massivevone\ and
later confirmed with independent calculations by \Soares.} are given by the equations
$B'_{\a\b}=\g^{mnp}_{\a\b}B'_{mnp}$ with $B'_{mnp}$ in \bpabc, $H'_{m\a}$ from \Hpdef,
$C^{'\b}{}_\a$ from \Cpab\ and $F'_{\a mn}$ from \Fpamn.
Dropping the $'$ superscript from the notation and using $k^m=k^m_1+k^m_2$, one can show using the
SYM equations of motion \SYMBRST\ that the
above expression satisfies \BCpaper
\eqnn\massBC
$$\eqalignno{
B_{\a\b} &= \g^{mnp}_{\a\b}B_{mnp}\,,&\massBC\cr
H_{m\a} &= {3\over7}(\g^{np})_\a{}^\b D_\b B_{mnp}\,,\cr
C^\b{}_\a &= \frac{1}{4}ik_m(\g^{mnpq})^\b_{\ \a}B_{npq}\,,\cr
F_{\a mn} &= \frac{1}{16} \Big(
		7ik_{[m}H_{n]\a}
		+ ik_q(\g_{q[m})_\a^{\ \b}H_{n]\b}
	\Big)\,,
}$$
where ($[mn]=mn-nm$)
\eqn\BmnpBC{
B_{mnp} =
{1\over9}\ap^2 (W_1\g^{k^1 k^2 mnp}W_2)
+\frac{1}{18}\ap^2\Big[ik^{2q}F^{1}_{q[m}F^{2}_{np]} +(1\leftrightarrow 2) \Big]\,,\quad 2\ap
k_1\cdot k_2 = -1\,.
}
Moreover, the
equations of motion of the superfields defined in \lfields\ are given by
\eqnn\BeomA
$$\eqalignno{
Q(\l B)_\a &= (\l\g^m)_\a (\l H)_m\,, &\BeomA\cr
Q(\l H^m) &= (\l\g^m C\l)\,,\cr
Q(C\l)^\a &= {1\over4}(\l\g^{mn})^\a (\l F)_{mn}\,,\cr
Q(\l F)_{mn} &= {1\over2}\p_{[m}(\l\g_{n]} C\l)
-{1\over16}\p^p(\l\g_{[m} C\g_{n]p}\l)\,,
}$$
and resemble the massless SYM equations of motion \SYMBRST. Note that the gauge transformations
\ghty\ preserve the first three equations but not the last. Finally, in the Berkovits-Chandia gauge
the unintegrated vertex operator at mass level one becomes
\eqn\Vmass{
V= [\l^\a[\p\t^\b B_{\a\b}]_0]_0 + [\l^\a[\Pi^m H^m_\a]_0]_0 + 2\ap [\l^\a[d_\b C^\b{}_\a]_0]_0
+ \ap[\l^\a[N^{mn} F_{\a mn}]_0]_0\,,
}
where the normal-ordering bracket $[AB]_0$ is defined in \NO, and we used \NlaABC. Alternatively,
it can also be suggestively rewritten in terms of the definition \lfields\ as
\eqn\VmassAg{
V= [\p\t^\a (B\l)_\a]_0 + [\Pi^m (\l H^m)]_0 + 2\ap [d_\a (C\l)^\a]_0
+ \ap[N^{mn} (\l F)_{mn}]_0\,,
}
resembling the massless integrated vertex operator \masslessU.

In the context of scattering amplitudes, the massless representation of massive superfields can be viewed as
swapping a string label, say ${\underline k}$ for the massive string, by a pair of massless string
labels, say $i,j$ ($i{=}1$ and $j{=}2$ in the example of \BmnpBC). At the level of superfields (both in the OPE
or Berkovits-Chandia gauge),
this swap will be denoted by
\eqn\swap{
{\underline k}\to i,j\,,\qquad 2\ap (k_i\cdot k_j) = -1\,,
}
where the constraint on the momenta must accompany the change of the superfields.

\newnewsec\ampsec Relation between massive and massless string amplitudes

In this section we will show that using the massless parameterization
of the massive superfields gives rise to an explicit
relation between massive and massless amplitudes of the open string. The combinatorics of this
problem can
be described by an algorithm \cdescent\ closely related to the descent algebra
\refs{\mackey,\garsiaReutenauer}. To see this relation, we
reinterpret the factorization condition to perform the equivalent of the sum over intermediate
polarizations on a single amplitude rather than a quadratic expression connected via a propagator.

\newsubsec\tptsec $3$-point massive amplitude revisited

The tree-level scattering amplitude of two massless and one massive state was firstly
computed using the pure spinor formalism in \PSthreemass. The result, despite correct, was obtained in a
rather convoluted way using the OPEs of the pure spinor formalism after performing the $\t$
expansions of the massive superfields obtained in \masstheta. Consequently, the simplicity of the result was lost.
However, one can
exploit the cohomological structure of the pure spinor formalism together with
the equations of motion \Beom\ to \Feom\ to obtain a simple answer written in pure spinor superspace.

We start with the three-point amplitude prescription
for massless strings labelled by $1$ and $2$ and one first-level massive string
labelled by ${\underline 3}$
\eqn\tptpresc{
4i\ap^2A(1,2|{\underline 3})= \langle V^{(0)}_1 V^{(0)}_2 V_{\underline3}^{(1)}\rangle
}
where $V^{(0)}$ and $V^{(1)}$ denote the massless and first-level massive unintegrated vertices
\masslessV\ and
\Vmass\ (the superscript indicates the mass and has been added for clarity).
The normalization on the left-hand side was
chosen to make the amplitude dimensionless, $[A(1,2|{\underline3})]=0$.

We will use the techniques of \refs{\fourtree,\fivetree,\nptTree} in which the OPEs among the
vertices are evaluated up to the plane-wave factors. This results in a chiral CFT correlator 
multiplying an overall Koba-Nielsen factor \PSthreemass
\eqn\KNt{
{\cal I} = |z_{12}|^{2\ap k_1\cdot k_2}|z_{13}|^{2\ap k_1\cdot k_3}|z_{23}|^{2\ap k_2\cdot k_3}=
{z_{13}z_{23}\over z_{12}}
}
which follows from momentum conservation and $k_1^2=k_2^2=0$, $k_3^2 = -1/\ap$.
In addition, the CFT correlator is evaluated up to BRST exact
terms and this will be indicated by $A\sim B$. Defining
\eqn\Lto{
L_{{\underline3}1} \sim [V_{\underline3}^{(1)}V^{(0)}_1]_1
}
we get
\eqn\tptatmp{
4i\ap^2A(1,2|{\underline 3}) = \Big({1\over z_{31}}\langle L_{{\underline3}1}V_2\rangle
- {1\over z_{32}}\langle V_1L_{\underline32}\rangle\Big) {\cal I} =
- {z_{23}\over z_{12}}\langle L_{\underline31}V_2\rangle
+ {z_{13}\over z_{12}}\langle V_1 L_{\underline32}\rangle\,.
}
A straightforward OPE calculation yields
\eqn\Ltotmp{
L_{\underline31} = -2\ap(\l H_3^m)\p_m V_1 + 2\ap (C_3\l)^\b D_\b V_1
-{\ap\over2}(\l F_3)_{mn}(\l\g^{mn}A_1)\,.
}
The equation of motion \SYMBRST\ can be used to rewrite the second term of \Ltotmp\ as
\eqnn\fgt
$$\eqalignno{
2\ap (C_3\l)^\b D_\b V_1 &= -2\ap (C_3\l)^\b QA_\b^1 + 2\ap(\l\g^m C_3\l)A_1^m &\fgt\cr
&= -2\ap Q(A_1C_3\l)+{\ap\over2}(\l F_3)_{mn}(\l\g^{mn}A_1) + 2\ap(\l\g^m C_3\l)A_1^m\,,
}$$
leading to cancellations in \Ltotmp.
Furthermore,
dropping the BRST exact term we arrive at
\eqnn\LtoA
$$\eqalignno{
L_{\underline31} &\sim - 2\ap (\l H_3^m)\p_m V_1 + 2\ap (\l\g^m C_3\l)A_1^m \cr
& \sim - 2\ap (\l H_3^m)QA^1_m + 2\ap(\l H_3^m)(\l\g_m W_1) + 2\ap (\l\g^m C_3\l)A_1^m \cr
& \sim 2\ap Q\big((\l H_3^m)A^1_m\big) + 2\ap(\l H_3^m)(\l\g_m W_1)\cr
& \sim 2\ap (\l H^m_3)(\l\g_m W_1)\,,&\LtoA
}$$
where we used the equations of motion \Heom\ and $QA^1_m = \p_m V_1 + (\l\g_m W_1)$.
It is easy to see that $QL_{\underline31}=0$, as expected from the definition \Lto\ and the
identities \jmassless\ and \QVmass.
After plugging in $L_{{\underline i}j}$ from \Lto\ into \tptatmp\
the three-point amplitude becomes
\eqn\tpta{
4i\ap^2 A(1,2|{\underline 3})=
- 2\ap {z_{23}\over z_{12}}\langle (\l\g_m W_1)V_2(\l H^m_3)\rangle
- 2\ap {z_{13}\over z_{12}}\langle V_1 (\l\g_m W_2)(\l H^m_3)\rangle
}
To simplify this answer further we will need the following:
\proclaim Lemma. In pure spinor superspace, the following is true:
\eqn\supproof{
\langle V_2(\l\g_m W_1)(\l H_3^m)\rangle = \langle V_1(\l\g_m W_2)(\l H_3^m)\rangle\,.
}

\noindent{\it Proof.}
Note that
\eqn\QcW{
Q(\l\g^m\cW_{12}) = V_1(\l\g^m W_2) -V_2(\l\g^m W_1)\,,
}
where $\cW_{12}$ is the
Berends-Giele current defined in \partIcohomology. Therefore,
\eqnn\ptmpI
$$\eqalignno{
\langle V_1(\l\g_m W_2)(\l H_3^m)\rangle &= \langle Q(\l\g_m \cW_{12})(\l H_3^m)\rangle
+ \langle V_2(\l\g_m W_1)(\l H_3^m)\rangle &\ptmpI\cr
&= \langle (\l\g_m \cW_{12})Q(\l H_3^m)\rangle + \langle V_2(\l\g_m W_1)(\l H_3^m)\rangle\cr
&= \langle (\l\g_m \cW_{12})(\l\g^m C_3\l)\rangle + \langle V_2(\l\g_m W_1)(\l H_3^m)\rangle\cr
&= \langle V_2(\l\g_m W_1)(\l H_3^m)\rangle
}$$
where we used BRST integration by parts to arrive at the second line followed by the equation of
motion \Heom\ and
the identity $(\l\g_m)_\a(\l\g^m)_\b = 0$ in the last line, finishing the proof.~\qed

Finally, using the Lemma \supproof\ in \tpta,
the three-point amplitude of two massless states and one first-level massive state becomes
\eqn\tptaf{
A(1,2|{\underline 3}) ={i\over2\ap} \langle V_1 (\l\g_m W_2)(\l H^m_3)\rangle\,.
}
This is
independent of worldsheet positions as expected from M\"obius invariance. Moreover, it is easy to show
that the amplitude is BRST invariant.

\newsubsec\tptmassless Massless representation of the massive $3$-point amplitude

Plugging in the massless representation of
the massive
superfield $(\l H_3^m)$ in the OPE gauge\foot{Using BRST cohomology manipulations, one can
show that the pure spinor superspace
expression \masslessrep\ is also obtained if  $(\l H_3^m)$ in the Berkovits-Chandia gauge is used
instead.} given in \lHmsimple\ (with the relabeling $1\rightarrow3$, $2\rightarrow4$)
into \tptaf\ one gets
\eqn\masslessrep{
A(1,2|{\underline3})\big|_{{\underline3}\to3,4} =
 \langle V_1(\l\g_m W_2)\bigl(F_3^{mn}k_4^n V_4
 + k_3^{m}(\l\g^n W_3)A^n_4\bigr)\rangle\,,
}
where ${\underline3}\to3,4$ defined in \swap\ represents the change to the massless representation
of the massive superfield $(\l H_3^m)$, and
we dropped the BRST exact term $k_3^m Q(W_3A_4)$ in $(\l H_3^m)$ from \lHmsimple\ because
$V_1(\l\g^m W_2)$ is BRST closed. Using the equation of motion $k_4^n V_4 = QA^n_4 - (\l\g^n W_4)$
one rewrites the first term inside the parenthesis in \masslessrep\ as
\eqn\tmpfirst{
F_3^{mn}k_4^n V_4 = Q(F_3^{mn} A_4^n) - F_3^{mn}(\l\g^n W_4) - k_3^{[m} (\l\g^{n]} W_3)A_4^n\,.
}
Therefore, discarding the BRST exact term from \tmpfirst, one obtains
\eqn\tmpsec{
F_3^{mn}k_4^n V_4
 + k_3^{m}(\l\g^n W_3)A^n_4 \sim - F_3^{mn}(\l\g^n W_4) + (\l\g^m W_3)(k_3\cdot A_4)\,.
}
Plugging \tmpsec\ into \masslessrep\ and using $(\l\g_m)_\a(\l\g^m)_\b = 0$ to drop the
second term from \tmpsec\ one finally arrives at
\eqn\finaltpt{
A(1,2|{\underline3})\big|_{{\underline3}\to3,4} =
 -\langle V_1(\l\g_m W_2) F_3^{mn}(\l\g_n W_4)\rangle\,.
}
The superspace expression in the right-hand side of \finaltpt\ is easily identified
as the kinematic factor of the massless four-point open-string amplitude at one loop
\refs{\MPS,\fourtree}
\eqn\foptone{
A(1,2,|{\underline3})\big|_{{\underline3}\to3,4} = - \langle C_{1|2,3,4}\rangle\,,\qquad 2\ap k_3\cdot k_4=-1\,,
}
where $C_{1|2,3,4}$ is the four-point BRST invariant defined in \refs{\oneloopbb,\EOMbbs}
whose explicit component expansion \PSteight\ can be downloaded from \website. When all states are bosonic,
$\langle C_{1|2,3,4}\rangle$ is proportional to the famous $t_8F^4$ combination,
where the $t_8$ tensor can be found in \GSteight.
The constraint in the momenta is written
explicitly for emphasis (as it is already implicitly required by \swap).
Alternatively,
the four-point BRST invariant
was shown to be proportional to
the $\ap^2$ string correction of the massless four-point
amplitude, denoted by $A^{F^4}(1,2,3,4)$ \oneloopbb
\eqn\trelAFq{
A(1,2|{\underline 3})\big|_{{\underline3}\to3,4} = -A^{F^4}(1,2,3,4)\,,\qquad 2\ap k_3\cdot k_4=-1\,.
}
Since the pure spinor superspace expression \finaltpt\
contains four superfields
and it is BRST invariant, one can expand it in components using regular four-point kinematics
and apply the massive kinematics constraint $2\ap(k_3\cdot k_4) = -1$ at the end of the calculations. Using the
normalization $\langle (\l\g^m\t)(\l\g^n\t)(\l\g^p\t)(\t\g_{mnp}\t)\rangle = 2880\ap^2$ of the pure
spinor bracket \psf\ we get
\eqn\tptcomp{
A(1,2|{\underline3})\big|_{{\underline3}\to3,4}
= -\ap^2 (k_1\cdot k_2)(k_2\cdot k_3)\AYM(1,2,3,4)\,,\qquad 2\ap k_3\cdot k_4=-1\,,
}
where $\AYM(1,2,3,4)$ represents the four-point SYM field-theory amplitude.

\paragraph{Component expansion} On the one hand, the component expansion of the amplitude \tptaf\ when all external
states are bosonic
can be computed using the theta expansion of $H^m_\a$ found in \masstheta,
\eqn\fof{
A(1,2|{\underline3}) = \ap f_1^{mp}f_2^{pn}g_{{\underline 3}mn} + 2ie^m_1 k_2^ne_2^p b_{{\underline 3}mnp}
+ (1\leftrightarrow2)\,,
}
where $g_{mn}$ is the symmetric traceless and $b_{mnp}$ is the $3$-form polarization
subject to $k^m g_{mn} = k^m b_{mnp} = 0$. The length dimensions of the quantities in \fof\ are
\eqn\lengthA{
[A(1,2|{\underline3})] = 0,\quad [g_{mn}] = 0,\quad [b_{mnp}]=1\,.
}
In addition, $f_1^{mn} = k^m_1 e^n_1 - k^n_1 e^m_1$ is
the component field-strength and we rescalled the overall normalization of the amplitude given in
\masstheta\ for convenience.

As shown in \masstheta, the massive polarizations of the open string can be extracted from
the massive superfields in the Berkovits-Chandia gauge as
\eqn\gmnbmnpdef{
g_{mn} = {1\over64}(D\g_{(m}H_{n)})\big|_{\t=0},\qquad
b_{mnp} = {9\over8}B_{mnp}\big|_{\t=0}\,,
}
where the overall normalizations were chosen for convenience. Using the massless representations of
the massive superfields $H^m_\a$ and $B_{mnp}$ from \massBC\ in the Berkovits-Chandia gauge with
labels $1$ and $2$ as in \BmnpBC\ yields \massivevone
\eqnn\masspol
$$\eqalignno{
g_{mn}(1,2) &= {1\over8}\big( e_1^m e_2^n
          + e_1^n e_2^m
          - {1\over3} \d_{mn} (e_1\cdot e_2)\big) &\masspol\cr
& + {1\over24}\ap  \big(
          2(k_1^m k_1^n
          - 2 k_1^m k_2^n) (e_1\cdot e_2)\cr
&\qquad\quad{}	  + 6 (k_2^m e_1^n
          + k_2^n e_1^m) (k_1\cdot e_2)
	  - \d_{mn} (k_1\cdot e_2) (k_2\cdot e_1)
+ (1\leftrightarrow2)
          \big)\cr
&       + {1\over6}\ap^2 k_{12}^m k_{12}^n(k_1\cdot e_2)(k_2\cdot e_1)\cr
b_{mnp}(1,2) &=
       {i\over16} \ap  \big(
       k_1^{[m}e_1^n e_2^{p]} + k_2^{[m}e_2^n e_1^{p]}
          \big)\cr
       &+ {i\over8}\ap^2 \big(
        k_1^{[m}k_2^n e_2^{p]}(k_2\cdot e_1)
 + k_2^{[m}k_1^n e_1^{p]}(k_1\cdot e_2)
%
          \big)\,,\qquad 2\ap (k_1\cdot k_2) = -1\,,
}$$
where the notation $g_{mn}(1,2)$ emphasizes the labels $1$ and $2$ of the massless polarizations
on the right-hand side.
Using the transversality $(k_i\cdot e_i)=0$ and that the states $1$ and $2$ are massless,
$k_1^2=k_2^2=0$, together with the
constraint $2\ap (k_1\cdot k_2) = -1$,
one can easily check that they are transverse, $k_{12}^m g_{mn}(1,2)=0$
and $k_{12}^m b_{mnp}(1,2)=0$ as well as traceless symmetric ($g_{mn}$) and antisymmetric ($b_{mnp}$)
in their vectorial indices.

Finally, it follows from the above discussion that
the two expressions \fof\ and \tptcomp\ of the three-point amplitude with one massive and
two massless states are related by
\eqn\polrel{
g_{{\underline3}mn}=g_{mn}(3,4)\,,\quad
b_{{\underline3}mnp}=b_{mnp}(3,4)\,,\qquad
2\ap (k_3\cdot k_4) = -1\,.
}
More generally, we define in analogy with \swap\ the same notation ${\underline k}\to i,j$ to
represent the swap to massless polarizations in place of the massive polarizations:
\eqn\Hmap{
{\underline k}\to i,j:\quad g_{{\underline k}mn}=g_{mn}(i,j)\,,\quad
b_{{\underline k}mnp}=b_{mnp}(i,j)\,,\qquad 2\ap (k_i\cdot k_j) = -1\,,
}
where $g_{mn}(i,j)$ and $b_{mnp}(i,j)$ are given by \masspol.

\newsubsubsec\simpsec Counting the degrees of freedom

The expressions in \masspol\ explicitly show how the massive polarizations are built out of the
massless polarizations. Recall that $g_{mn}$ transform as the rank-two symmetric traceless tensor
of $SO(9)$ with $44$ degrees of freedom and
$b_{mnp}$ as
rank-three totally antisymmetric tensor of $SO(9)$ containing $84$ degrees of freedom.
The presence of massive excitations in the OPE guarantees that this construction is correct,
but one may wonder how the counting
of degrees of freedom works out as $e^m_i$ represents the $8$ of $SO(8)$. As we will see, it is
crucial that the massive polarizations are formed using not only the massless polarizations but also
combinations of momenta. To see this, it is convenient to define
\eqnn\polredefs
$$\eqalignno{
\zeta^m_1 &= e^m_1 + 2\ap(k_2\cdot e_1)k_1^m, &\polredefs\cr
\zeta^m_2 &= e^m_2 + 2\ap(k_1\cdot e_2)k_2^m,
}$$
that satisfy $(\zeta_1\cdot k_{12}) = (\zeta_2\cdot k_{12}) = 0$ due to $2\ap(k_1\cdot k_2) = -1$,
where $k_{12}^m =k_1^m+k_2^m$. Because of this, both $\zeta^m_1$ and $\zeta^m_2$ are vectors of
$SO(9)$. Finally, one can show that
\masspol\ can be written as
\eqnn\masspolson
$$\eqalignno{
g_{mn}(1,2) & = {1\over8}\big(\zeta_1^m\zeta_2^n+\zeta_1^n\zeta_2^m\big)
-{1\over24}(\zeta_1\cdot\zeta_2)\big(\d^{mn}-2\ap k^{mn}_{12}\big), &\masspolson\cr
b_{mnp}(1,2) &= {i\over16}k_{[12]}^{[m}\zeta_1^n\zeta_2^{p]}
}$$
where $k_{[12]}^m = k_1^m - k_2^m$ and $k_{12}^{mn} = k_1^m k_1^n + k_2^m k_2^n - 2k_1^m k_2^n -
2k_1^n k_2^m$ satisfy $(k_{[12]}\cdot k_{[12]}) = 1/\ap$ and $k_{12}^{mm} = 2/\ap$.
This shows that $g_{mn}(1,2)$ is indeed a symmetric and traceless rank-two tensor of $SO(9)$ of
dimension $44$ while
$b_{mnp}(1,2)$ is a rank-three totally antisymmetric tensor of $SO(9)$ of dimension $84$.

\newsubsec\mmsec Massive amplitudes as linear combinations of massless amplitudes

The observation \foptone\ generalizes to higher multiplicities. To see this, we use the
perturbiner construction of $A(P|{\underline n})$
recently found in \massSweden.
In that paper, the superstring amplitude involving $n{-}1$ massless states
and one massive state $\underline{n}$ was packaged in terms of $(n{-}3)!$
worldsheet integrals $F^P_Q$ and partial subamplitudes $A(1,P,n{-}1|\underline{n})$ as
\eqn\stringmassS{
{\cal A}(1,Q,n{-}1,\underline{n}) = \sum_{P \in S_{n-3}} F^P_Q
A(1,P,n{-}1|\underline{n})\,,
}
where $P$ and $Q$ are words comprised of particle labels (letters) and $F^P_Q$
have the same functional form as the string disk integrals
in the massless string scattering amplitude \refs{\nptTree,\nptTreeII,\drinfeld,\BGap};
the only difference stems from the massive constraint $k_{\underline{n}}^2 =
-1/\ap$ affecting the relations among Mandelstam variables.
When all external states are bosonic, the partial amplitudes $A(P|\underline{n})$
with $|P|=n{-}1$ massless states and one massive state at the first massive
level are given by \massSweden,
\eqn\APun{
A(P|{\underline n}) = \phi_P^{mn} g_{{\underline n}\,mn} + \phi_P^{mnp} b_{{\underline n}\,mnp}\,,
}
where the $|P|$ massless states are encoded in
\eqnn\allmul
$$\eqalignno{
\phi_P^{mn} &= \ap \sum_{XY=P} f_X^{ma}f_Y^{an} + {\rm cyc}(P)\,, &\allmul\cr
\phi_P^{mnp} &= 2i\sum_{XY=P} e_X^m k_Y^n e_Y^p - {4i\over3}\sum_{XYZ=P} e_X^m e_Y^n e_Z^p + {\rm
cyc}(P)\,,
}$$
where $e_X^m$ and $f_X^{mn}$ are the Berends-Giele multiparticle polarizations of \BGBCJ,
\eqnn\multSYM
$$\eqalignno{
e_P^m &= {1\over k_P^2}\sum_{XY=P}\bigl[ e^m_Y(k_Y\cdot e_X) + f_X^{mn}e_Y^n - (X\leftrightarrow Y)\bigr]\,, &\multSYM\cr
f_P^{mn} &= k_P^m e_P^n - k_P^n e_P^m - \sum_{XY=P}\big(e_X^m e_Y^n - e_X^n e_Y^m\big)\,.
}$$
In addition,
the notation ${}{+}{\rm cyc}(P)$ instructs to add the cyclic permutations
of the letters in $P$, $XY{=}P$ denotes the deconcatenations of
$P$ into non-empty words $X$ and $Y$, and $k_{ij \ldots p}^m = k^m_i + k^m_j + \cdots + k^m_p$.

Upon plugging the massless representations \masspol\ of the massive polarizations into the
amplitude \APun, replacing the massive state ${\underline n}$ by two massless states labelled $n$
and $n{+}1$, straightforward but tedious calculations\foot{We acknowledge the use of {\tt FORM}
\FORM.} show that:
\eqnn\affs
$$\eqalignno{
A(1,2|{\underline3})\big|_{{\underline3}\to 3,4} &= -\langle C_{1|2,3,4}\rangle\,,\qquad 2\ap k_3\cdot k_4=-1\,, &\affs\cr
A(1,2,3|{\underline4})\big|_{{\underline4}\to 4,5} &= -\langle C_{1|23,4,5}\rangle\,,\qquad 2\ap k_4\cdot k_5=-1\,,\cr
A(1,2,3,4|{\underline5})\big|_{{\underline5}\to 5,6} &= -\langle C_{1|234,5,6}\rangle\,,\qquad 2\ap k_5\cdot k_6=-1\,,
}$$
where the substitution rule in the left-hand side is given by \Hmap, and the constraint in the
momenta is written down for emphasis (in addition to featuring in \Hmap).
The explicit components of the scalar BRST invariants are available to download from \website\ and
can be used to check the relations above.
These results suggest the following generalization,
\eqn\relmm{
A(1,P|{\underline n})\big|_{{\underline n}\to n,n+1} = -\langle C_{1|P,n,n{+}1}\rangle\,,\qquad 2\ap k_n\cdot k_{n+1}=-1\,,
}
relating the massive string amplitude with one massive external state to the $\ap^2$ sector
of the massless tree-level string amplitudes.

\paragraph{Massless string amplitudes at $\ap^2$ order}
To make the connection to the $\ap^2$ correction of massless string tree amplitudes even
clearer, we can explicitly rewrite \relmm\ in terms of $A^{F^4}$, the $\ap^2$ corrections to
massless string amplitudes defined by \oneloopbb
\eqn\mzerostr{
A(1,2, \ldots,n) = A^{\rm YM}(1,2, \ldots,n) + \ap^2\zeta_2 A^{F^4}(1,2, \ldots,n) + {\cal
O}(\ap^3)\,.
}
To see this, we note that the BRST invariants $\langle C_{1|X,Y,Z}\rangle$
can be expanded in terms of permutations of $A^{F^4}$, as argued in \oneloopbb.
The precise permutations in this expansion turns out to be related to the Solomon descent algebra,
as described in \cdescent. In particular, for each $n$-point BRST invariant $C_{1|X,Y,Z}$ characterized by words
$X,Y$ and $Z$, one can define precise permutations $\g_{1|X,Y,Z}$
of $n$ labels (from the letters in $1,X,Y,Z$) dubbed {\it BRST-invariant
permutations}. For example,
\eqn\gamperms{
\g_{1|2,3,4} =   W_{1234}
          + W_{1243}
          + W_{1324}
          + W_{1342}
          + W_{1423}
	  + W_{1432}\,,
}
where a permutation $\s$ is written as $W_\s$ for typographical reasons.

The relation found in \cdescent\ expands the BRST invariants as permutations of $A^{F^4}$,
\eqn\bperms{
\langle C_{1|X,Y,Z}\rangle = {1\over6}A^{F^4}(\g_{1|X,Y,Z})\equiv {1\over
6}\sum_{\s\in\g_{1|X,Y,Z}}A^{F^4}(\s)\,.
}
For instance, using the permutations in \gamperms\ one gets
\eqnn\CfourAfour
$$\eqalignno{
\langle C_{1|2,3,4}\rangle &= {1\over6}A^{F^4}(\g_{1|2,3,4}) &\CfourAfour\cr
&= {1\over6}\big( A^{F^4}(1234)
          + A^{F^4}(1243)
          + A^{F^4}(1324)\cr&\qquad{}
          + A^{F^4}(1342)
          + A^{F^4}(1423)
	  + A^{F^4}(1432)\big)\cr
&= A^{F^4}(1234)\,,
}$$
where in the last line we used the total symmetry of $A^{F^4}(1234)$.

In view of the identity \bperms, the general observation \relmm\ yields a expansion in permutations of $A^{F^4}$
\eqn\CtoAFq{
A(1,P|{\underline n})\big|_{{\underline n}\to n,n+1}
= -{1\over6}A^{F^4}(\g_{1|P,n,n{+}1}),\qquad 2\ap k_n\cdot k_{n+1} = -1\,.
}
For example, one gets
\eqnn\fptAFq
$$\eqalignno{
-6A(1,2,3|{\underline 4})\big|_{{\underline4}\to4,5} &=
       \AFq(1,2,3,4,5)
       - \AFq(1,2,3,5,4)
       - \AFq(1,2,4,3,5)
       + \AFq(1,2,4,5,3)
       + \AFq(1,2,5,3,4)
       - \AFq(1,2,5,4,3) &\fptAFq\cr
&       + \AFq(1,3,2,4,5)
       - \AFq(1,3,2,5,4)
       + \AFq(1,3,4,2,5)
       - \AFq(1,3,5,2,4)
       - \AFq(1,4,2,3,5)
       - \AFq(1,4,3,2,5)\,,
}$$
where the explicit permutations in $\g_{1|23,4,5}$ and the algorithm to generate them can be found
in \cdescent. Note that we used the parity and cyclicity in the form of $A^{F^4}(1,2, \ldots,n)=(-1)^n
A^{F^4}(1,n,n-1, \ldots,2)$ to reduce the number of terms in \fptAFq.

\newsubsec\facsec Relating massive and massless string amplitudes via unitarity

In this section we will show that the result \relmm\ can be explained from the factorization
of the massless string amplitude in its first massive pole\foot{See \nunez\ for similar
considerations with the RNS formalism.} . To see this, one
computes the residue of the massless $n$-point tree-level amplitude when
$s_{n{-}1,n} = -1$ (setting $\ap=\half$).

\newsubsubsec\fptfact $4$-point factorization

The massless $4$-point tree amplitude is given by
\eqn\fpttree{
{\cal A}_4 = {\cal A}(1,2,3,4) = \langle C_{1|2,3,4}\rangle
{\Gamma(s_{34})\Gamma(s_{23})\over\Gamma(1+s_{34}+s_{23})}\,.
}
Using the well known result that $\res{x=-n}(\Gamma(x))=(-1)^n/n!$ we get \nimaLorenzYuSebas
\eqn\Cfromfac{
\res{s_{34}=-1}({\cal A}_4)=-\langle C_{1|2,3,4}\rangle\,,\quad s_{34}=-1\,,
}
which explains the first line of \affs.

\newsubsubsec\fiptfact $5$-point factorization

The five-point analysis proceeds similarly. But note that the M\"obius symmetry
gauge fixing $(z_1,z_4,z_5)\to(0,1,\infty)$ in the usual formula \refs{\nptTree,\nptTreeII}
\eqn\fivetreeST{
{\cal A}_5={\cal A}(1,2,3,4,5) = A^{\rm YM}(12345)F_{23}+ A^{\rm YM}(13245)F_{32}
}
with
\eqn\Fints{
F_{23} = \int_0^1 dz_3 \int_{0}^{z_3} dz_2\,  {s_{12}s_{34}\over z_{12}z_{34}}\,{\cal I}_5, \quad
F_{32} = \int_0^1 dz_3 \int_{0}^{z_3} dz_2\,  {s_{13}s_{24}\over
z_{13}z_{24}}\,{\cal I}_5
}
is not well suited to obtain the residues with respect to $s_{45}$, since
$z_5\to\infty$ and the corresponding factor of $z_{54}^{s_{45}}$ is absent in
the Koba-Nielsen factor
${\cal I}_5 =
z_{21}^{s_{12}}z_{31}^{s_{13}}z_{32}^{s_{23}}z_{42}^{s_{24}}z_{43}^{s_{34}}\,.
$
However, one can exploit the cyclicity of ${\cal A}_5$ to compute the residue
as $s_{12}=-1$ and then apply three cyclic rotations in succession
$s_{12}\to s_{23} \to s_{34} \to s_{45}$ to obtain the residue as $s_{45}=-1$.
The calculations done in \bianchifac\ show that
\eqn\residu{
\res{s_{12}=-1}({\cal A}_5) =
s_{34} B(s_{13}+s_{23},s_{34})\Big(
A^{\rm YM}(12345)\Bigl(s_{23}-{s_{24}(s_{13}+s_{23})\over s_{35}}\Bigr)
-A^{\rm YM}(13245){s_{13}s_{24}\over s_{35}}\Big)
}
where $B(x,y) = {\Gamma(x)\Gamma(y)\over\Gamma(x+y)}$ is the Beta function. After noticing that
\eqn\frombia{
-{1\over s_{12}}\langle C_{5|43,2,1}\rangle =
A^{\rm YM}(12345)\Bigl(s_{23}-{s_{24}(s_{13}+s_{23})\over s_{35}}\Bigr)
-A^{\rm YM}(13245){s_{13}s_{24}\over s_{35}}\,,
}
which can be shown
using the algorithm of \EOMbbs\ to rewrite
\eqn\ctoaym{
\langle C_{5|43,2,1}\rangle =
 A^{\rm YM}(52134) s_{12} s_{13}
        - A^{\rm YM}(52143) s_{12} s_{14}
}
and expressing the result in the basis of $A^{\rm YM}(12345)$ and $A^{\rm YM}(13245)$, one gets
\eqn\residuagain{
\res{s_{12}=-1}\bigl({\cal A}_5\bigr) =
s_{34} B(s_{13}+s_{23},s_{34})
\langle C_{5|43,2,1}\rangle\,,\qquad s_{12}=-1\,.
}
The cyclic rotations of \residuagain\ give rise to
\eqnn\trures
$$\eqalignno{
\res{s_{45}=-1}\bigl({\cal A}_5\bigr)
&= s_{12} B(s_{14}+s_{15},
s_{12})\langle C_{3|21,5,4}\rangle\,,\qquad s_{45}=-1 \cr
&= - s_{12} B(-s_{12}-s_{13},s_{12})
\langle C_{1|23,4,5}\rangle\,,\qquad s_{45}=-1\,,&\trures
}$$
where in the last line we used the canonicalization identity \partIcohomology\
$\langle C_{3|21,5,4}\rangle = -\langle C_{1|23,4,5}\rangle$ and momentum
conservation. This is compatible with the factorization
\eqn\fact{
\res{s_{45}=-1}\bigl({\cal A}_5\bigr) =
\sum_x {\cal A}(1,2,3|{\underline x}){\cal A}(4,5|{\underline x})\,,
}
where $\sum_x$ denotes a sum over the massive polarizations at the first mass level.
To see this, note that the three- and four-point string amplitudes with one
massive state $x$ are \massSweden\
($1+s_{23}=-s_{12}-s_{23}$)
\eqnn\tmamp
$$\eqalignno{
{\cal A}(1,2,3|{\underline x}) &= s_{12}B(-s_{12}-s_{13},s_{12}) A(1,2,3|{\underline x})\,, &\tmamp\cr
{\cal A}(4,5|{\underline x}) &= A(4,5|{\underline x})\,,
}$$
where $A(P|{\underline x})$ is given by \APun\ (see also \refs{\stielust,\olihigh}).
Compatibility of \trures, \fact\ and \tmamp\ requires that
\eqn\sumpo{
\sum_x A(1,2,3|{\underline x})A({\underline x}|4,5) = -\langle
C_{1|23,4,5}\rangle,\quad s_{45}=-1\,,
}
which can be explicitly checked using \sumovermassive\ below.

\newsubsubsec\sumPolsec Sum over intermediate massive polarizations

The justification given above for the first two lines of \affs\ was
obtained by computing the first massive residue of the massless
amplitudes. A more direct derivation follows from the interpretation
that the left-hand sides of \affs\ are given by an sum over intermediate massive polarizations
\eqn\interp{
A(1,2, \ldots,n-1|{\underline n})\big|_{{\underline n}\to n,n+1} =
\sum_x A(1,2, \ldots,n-1|{\underline x}) A(n,n+1|{\underline x})\,.
}
To see this, note that the massless representation rules of \masspol\ encapsulated in \Hmap\ 
follow from the factorization relations
\eqn\sumgb{
\sum_x g_{mn}(k)g_{pq}(-k)\phi^{pq}_{12} = g_{mn}(1,2),\quad
\sum_x b_{mnp}(k)b_{qrs}(-k)\phi^{qrs}_{12} = b_{mnp}(1,2)\,,
}
where $\phi^{pq \ldots}_P$ was defined in \allmul, and the momenta is $k=k_1+k_2$. Moreover,
the sum over the massive states $x$ is performed by the completeness relations \facrevisited,
\eqnn\sumovermassive
$$\eqalignno{
\sum_x g_{mn}(k)g_{pq}(-k)&={1\over 64}\Big((k_m k_p + 2\eta_{mp})(k_n k_q + 2\eta_{nq})
&\sumovermassive\cr
&\qquad{}-{1\over9}(k_m k_n + 2\eta_{mn})(k_p k_q + 2\eta_{pq}) + (m\leftrightarrow n)\Big)\cr
\sum_x b_{mnp}(k)b_{qrs}(-k)&={1\over
256}(k_mk_q+2\eta_{mq})(k_nk_r+2\eta_{nr})(k_pk_s+2\eta_{ps})+[mnp]\,,
}$$
where $[mnp]$ instructs to antisymmetrize over the indices $mnp$ and we set $\ap=\half$.

With this interpretation, the relations \affs\ can be written as
\eqnn\affsfac
$$\eqalignno{
\sum_x A(1,2|{\underline x})A(3,4|{\underline x}) &= -\langle C_{1|2,3,4}\rangle\,,\quad
2\ap(k_3\cdot k_4)= -1\,,&\affsfac\cr
\sum_x A(1,2,3|{\underline x})A(4,5|{\underline x}) &= -\langle C_{1|23,4,5}\rangle\,,\quad
2\ap(k_4\cdot k_5)= -1\,,\cr
\sum_x A(1,2,3,4|{\underline x})A(5,6|{\underline x}) &= -\langle C_{1|234,5,6}\rangle\,,\quad 2\ap(k_5\cdot k_6)= -1\,,
}$$
and have been explicitly verified. They suggest the generalization
\eqn\affsgen{
\sum_x A(1, P|{\underline x})A(n,n+1|{\underline x}) = -\langle C_{1|P,n,n+1}\rangle,\quad
2\ap(k_n\cdot k_{n+1})= -1\,,
}
as the equivalent statement to \relmm. However, note the interpretation
difference in how unitarity is actually implemented to arrive at the equivalent
results \affsgen\ and \relmm.

\newnewsec\consec Conclusions

In this paper we found an explicit realization of the massive superfields describing the open
string states at the first mass level in terms of massless SYM fields.
This was achieved through the calculation of OPEs between massless
vertices, giving rise to a massless representation in the so-called OPE gauge \Balbe-\Famn.
Additional manipulations were used to fix the gauge invariance of the unintegrated vertex operator
due to BRST-exact pieces, with the end result being the massless representation in the
Berkovits-Chandia gauge \massBC\ and \BmnpBC.

After simplifying the three-point amplitude of two
massless and one massive state obtained in \PSthreemass\ to a single pure spinor superspace
expression,
\eqn\tptafConc{
A(1,2|{\underline 3}) ={i\over2\ap} \langle V_1 (\l\g_m W_2)(\l H^m_3)\rangle\,,
}
the massless representation of the massive superfield $H^m_\a$
was then used through the superspace substitution \swap.
The resulting expression \finaltpt\ related, at the superspace level, the massive amplitude
\tptafConc\ to the $\ap^2$ correction
of the massless four-point open string amplitude as captured by the scalar BRST
invariants \partIcohomology. The generalization of this relation was proposed as
\eqn\relmmConc{
A(1,P|{\underline n})\big|_{{\underline n}\to n,n+1} = -\langle C_{1|P,n,n{+}1}\rangle\,,\qquad 2\ap k_n\cdot k_{n+1}=-1\,,
}
where the restriction ${\underline n}\to n,n{+}1$ on the left-hand
side was defined in \Hmap\ from superfield considerations
and later justified via factorization in \sumgb. The proposal \relmmConc\
was then explicitly checked in terms of polarizations and momenta up to ${\underline n}=6$ and
shown to be compatible with
unitarity via the residue of the massless amplitudes at their first massive pole. The translation
from the right-hand side of \relmmConc\ to linear combinations of $\ap^2$ massless amplitudes
$A^{F^4}$ (defined in \oneloopbb) follows from the descent algebra algorithm described in \cdescent.

It would be interesting to invert the relation \relmmConc\ to find the pure spinor superspace
expression of the partial massive amplitudes \APun. In addition, it may be possible
to turn the observations in this paper
into a constructive algorithm to compute $n$-point massive amplitudes (one first-level massive state
and $n{-}1$ massless states) starting from the known expressions of
the $\ap^2$ massless open string amplitudes\foot{Both of these points have been solved in
\towardsmass.}. Higher $\ap$ corrections of the massless amplitudes are expected to contain information about
massive amplitudes of a combination of states from higher mass levels. The references
\refs{\refo,\reft} may be helpful, and working out the precise
relations is left for future work.

\bigskip \noindent{\bf Acknowledgements:} We thank an anonymous referee for comments on the
companion paper \massivevone.
CRM was supported by a University Research
Fellowship from the Royal Society during the initial stages of this project.
MV is supported in part by the ``Young Faculty
Research Seed Grant Scheme'' of IIT Indore.
During the initial
stages of this work, LAY was supported by CRM's Royal Society University
Research Fellowship.

\appendix{A}{OPEs of non-free fields}
\applab\opeapp

\noindent As explained in \yellowbook, a free field is defined
as a field whose OPE with itself or its derivatives contain a single
{\it constant} term.
In the pure spinor formalism, the fields are not
in general free as can be seen from the OPE of $d_\a(z)d_\b(w)$ or
$N^{mn}(z)N^{pq}(w)$. In this case, the definition of normal ordering
of operators and the calculation of OPEs with normal ordered operators
is done following a generalization of the conventional Wick theorem rules, see e.g.
\refs{\bais,\yellowbook}.
Let us briefly review the calculation of OPEs involving composite operators
following the exposition of
\refs{\farril,\thielemans} (see also \refs{\axioms,\japwick}).

\newsubsec\OPEsec Operator product expansion of composite operators

The OPE of $A$ and $B$
is defined as ($N$ is a finite positive integer)
\eqn\opeAB{
A(z)B(w) = \sum_{n=-\infty}^{N} {[AB]_n(w)\over (z-w)^n}
}
and the normal-ordered product of $A$ and $B$, denoted $(AB)(w)$,
is given by
\eqn\NO{
(AB)(w) = \oint {dz\over z-w} A(z)B(w) = [AB]_0(w).
}
\paragraph{Generalized Wick theorem}
The calculation of nested OPEs of non-free fields can be
done entirely at the level of the OPE brackets introduced above.
The underlying techniques follow from the Borcherds identity
\eqnn\borcid
$$\displaylines{
\sum_{j=0}^\infty {p\choose j}[[AB]_{r+j+1}C]_{p+q+1-j} = \hfil\borcid\hfilneg\cr
\sum_{j=0}^\infty (-1)^j{r\choose j}\Bigl([A[BC]_{q+1+j}]_{p+r+1-j} -
(-1)^{r+ab}[B[AC]_{p+1+j}]_{q+r+1-j}\Bigr),\quad p,q,r\in{\Bbb Z}
}$$
which plays a major role in vertex operator algebra \borcherds. In the above, $a,b$ denote the Grassman
parities of $A$ and $B$, respectively.
The
two special cases of the Borcherds identity that are frequently used follow from
$(p{+}1{=}m,q{+}1{=}n,r{=}0)$ and $(p{=}0,q{+}1{=}n,r{+}1{=}m)$; they give rise
to identities for $[A[BC]_n]_m$ and $[[AB]_mC]_n$ \refs{\farril, \thielemans}:
\eqnn\ABCt
\eqnn\ABC
$$\eqalignno{
[A[BC]_n]_m &= (-1)^{ab}[B[AC]_m]_n +
\sum_{j=0}^{m-1}{m-1\choose j}[[AB]_{m-j}C]_{n+j}\,,\quad m\ge 1&\ABCt\cr
[[AB]_m C]_n &= \sum_{j=0}^\infty (-1)^j{m-1\choose j}\bigl(
[A[BC]_{n+j}]_{m-j}+(-1)^{m+ab}[B[AC]_{j+1}]_{m+n-j-1}\bigr)\qquad{} &\ABC
}$$
where we used ${m-1\choose j}={m-1\choose m-1-j}$ and relabeled
$m-1-j\to j$ in \ABCt.
In particular, when the composite operators are normal ordered
we get\foot{Note ${n\choose m}=(-1)^m{-n+m-1\choose
m}$ when $n<0$ implies ${-1\choose j}=(-1)^j$.}
\refs{\bais,\japwick}
\eqnn\frombais
\eqnn\ABCNO
$$\eqalignno{
[A[BC]_0]_n &= (-1)^{ab}[B[AC]_n]_0 + [[AB]_nC]_0
+ \sum_{i=1}^{n-1}{n-1\choose i}[[AB]_{n-i}C]_i &\frombais\cr
&=(-1)^{ab}\Bigl(
[B[AC]_n]_0 + \sum_{j=0}^\infty {(-1)^{j+n}\over j!}[\p^j[BA]_{j+n} C]_0
+ \sum_{i=1}^{n-1}(-1)^i[[BA]_i C]_{n-i}
\Bigr),\cr
[[AB]_0 C]_n &=\sum_{j=0}^\infty\bigl(
[A[BC]_{n+j}]_{-j} + (-1)^{ab}[B[AC]_{j+1}]_{n-j-1}
\bigr)&\ABCNO\cr
&=(-1)^{ab}\sum_{i=1}^{n-1}[B[AC]_{n-i}]_i 
+\sum_{j=0}{1\over j!}\Bigl(
[\p^j A[BC]_{n+j}]_0
+(-1)^{ab}[\p^j B[AC]_{n+j}]_0
\Bigr)
}$$
Repeated application of these rules allow the computation of OPE brackets with arbitrary nesting.
Some useful relations obeyed by the brackets are
\eqnn\ABCNONO
\eqnn\thielid
$$\eqalignno{
[A[BC]_0]_0 &= (-1)^{ab}[B[AC]_0]_0 + \sum_{i=1}^\infty (-1)^{1+i}{1\over i!}[\p^i[AB]_iC]_0
&\ABCNONO\cr
[[AB]_0 C]_0 &= [A[BC]_0]_0 + \sum_{i=1}{1\over i!}\bigl(
[\p^i A[BC]_i]_0
+ (-1)^{ab}[\p^i B[AC]_i]_0
\bigr) &\thielid\cr
}$$
and (with $n$ a non-negative integer):
\eqnn\ABn
\eqnn\pAB
\eqnn\ApB
\eqnn\ABderiv
\eqnn\ABderivi
$$\eqalignno{
[BA]_n &= (-1)^{n+{ab}}\Big([AB]_n + \sum_{i=1}^\infty (-1)^{i}{1\over i!}\p^i[AB]_{n+i}\Big) &\ABn\cr
[\p A B]_n &= (1-n)[AB]_{n-1} &\pAB\cr
[A\p B]_n &= \p[AB]_n + (n-1)[AB]_{n-1} &\ApB\cr
[AB]_{-n} &= {1\over n!}[\p^n A B]_0 &\ABderiv\cr
[AB]_{n-i}&={(-1)^i\over {n-1\choose i}}{1\over i!}[\p^i A B]_n &\ABderivi
}$$
Note that $\p[AB]_n = [\p A B]_n + [A\p B]_n$.
In addition, $[A,\;]_1$ is a graded
derivation over all other brackets. This means that
\eqn\deriv{
[A[BC]_n]_1 = [[AB]_1C]_n + (-1)^{ab}[B[AC]_1]_n\,.
}
Furthermore, if the conformal weights of $A$ and $B$ are $h_A$ and $h_B$ then
$[AB]_n$ has conformal weight
$h_A+h_B-n$, i.e., the bracket $[\;]_n$ has conformal weight ${-}n$.


\newsubsubsec\planesec OPEs of superfields in a plane-wave basis

A superfield $K_i\in[A_\a, A_m, W^\a,
F^{mn}]$ in a plane wave basis is expanded as
\eqn\Kplane{
K_i(z) = K_i(\t,X) = K_i(\t(z))e^{ik_i\cdot X(z)}\,,
}
for example $A^1_m(z) = A^1_m(\t(z))e^{ik_1\cdot X(z)}$.
We are interested in the OPE of two such superfields.
The definition of the OPE given in \opeAB\ needs to be generalized when the
operators involve plane-wave factors $e^{i k\cdot X}$ as the behavior
$$\eqalignno{
{:}e^{ik_1X(z)}{:}{:}e^{ik_2X(w)}{:} &=
(z-w)^{2\ap  k_1\cdot k_2}\bigl[1+(z-w)ik_1\cdot\p X(z) + {\cal
O}((z-w)^2)\bigr]{:}e^{ik_3\cdot X(w)}{:}
}$$
where $k_3=k_1+k_2$ is not of the form \opeAB\ unless $2\ap k_1\cdot k_2$ is
an integer. However, when $2\ap k_1\cdot k_2 = -1$ the
OPE can be written as
\eqn\KKOPE{
K_1(z)K_2(w) = \sum_{n=-\infty}^N{[K_1K_2]_n\over (z-w)^n}\,,\qquad 2\ap k_1\cdot k_2 = -1
}
with
\eqnn\charac
$$\eqalignno{
[K_1 K_2]_0(w) &=\p K_1(w) K_2(w)\,,&\charac\cr
[K_1 K_2]_1(w) &=K_1(w) K_2(w)\,,\cr
[K_1 K_2]_{n\ge2}(w) &= 0\,.
}$$
To see this, note
that there is no worldsheet singularity between the factors $K_i(\t)$; the OPE
singularity
comes entirely from the plane waves using $2\ap k_1\cdot k_2 = -1$,
\eqn\plOPE{
{:}e^{ik_1X(z)}{:}{:}e^{ik_2X(w)}{:} =
{e^{ik_1\cdot X(w)}e^{ik_2\cdot X(w)}\over (z-w)}\bigl[1+(z-w)ik_1\cdot\p X(z) + {\cal
O}((z-w)^2)\bigr]
}
The factors $K_i(\t)$ contribute
via the
Taylor expansion
\eqn\tayK{
K_1(\t(z))K_2(\t(w)) =K_1(\t(w))K_2(\t(w)) + (z-w)\p\t^\a \p_\a K_1(\t(w))
K_2(\t(w)) + {\cal O}((z-w)^2).
}
From \plOPE\ and \tayK\ it follows that
\eqnn\brac
$$\eqalignno{
[K_1 K_2]_0(w) &= (ik_1\cdot \p X K_1(\t(w))  + \p\t^\a \p_\a
K_1(\t(w))\bigr)e^{ik_1\cdot X(w)}
K_2(\t(w))e^{ik_2\cdot X(w)}\cr
&=\p K_1(w) K_2(w)\,. &\brac
}$$
Similarly,
$[K_1 K_2]_1(w) = K_1(w)K_2(w)$ and $[K_1 K_2]_{n\ge2}=0$.

\newsubsec\OPEsPS OPEs in the pure spinor formalism

Using conventions for the open string, some of the basic OPEs of the pure spinor
formalism used in this work are listed below (for brevity, the dependence on $w$ is omitted on the
right-hand side):
\eqnn\allopes
$$\displaylines{
\p\t^\a(z)\big\{\p\t^\b(w),\Pi^m(w), N^{mn}(w)\big\} \sim{\rm regular},\qquad
d_\a(z)\p\t^\b(w) \rightarrow  {\d^\b_\a\over (z- w)^2},\hfil\allopes\hfilneg\cr
d_\a(z) K(w) \rightarrow  {D_\a K\over z- w}, \quad
\Pi^m(z) K(w) \rightarrow - 2\ap {\p^m K\over z- w}, \quad
d_\a(z) \Pi^m(w) \rightarrow  {(\g^m\p\t)_\a\over z- w}
\cr
d_\a(z)d_\b(w) \rightarrow -  {1\over 2\ap}{\g^m_{\a\b}\Pi_m\over z- w}, \quad
\Pi^m(z)\Pi^n(w) \rightarrow -  2\ap{\eta^{mn}\over (z - w)^2}, \quad
d_\a(z)\t^\b(w) \rightarrow  {\d^\b_\a\over z- w}
\cr
J(z)J(w)\rightarrow - {4\over (z-w)^2},\quad
J(z)\l^\a(w)\rightarrow  {\l^\a\over z-w},\quad
N^{mn}(z)\l^\a(w) \rightarrow \half {(\g^{mn}\l)^\a\over z- w}
\cr
N^{mn}(z)N^{pq}(w)
         \rightarrow  {\delta^{p[m} N^{n]q}  - \delta^{q[m} N^{n]p}\over z-w}
 -3{\delta^{m[q}\delta^{p]n}\over(z-w)^2}
}$$
where $K(w)$ is a generic $10$D superfield that does not depend on derivatives $\p^k X^m$ and
$\p^k\t^\a$ with $k\ge1$.

\newsubsec\detailsapp Rearranging normal ordered brackets

The direct evaluation of the bracket $I_4\equiv \ap[[N^{mn}F^{mn}_1]_0[\l^\b A_\b^2]_0]_1$ using
the rules in \ABC\ and \ABCt\ gives
\eqn\prob{
I_4 = \ap [\l^\b[N^{mn}(F^{mn}_1 A^2_\b)]_0]_0
+{\ap\over2}(\g^{mn})^\b{}_\g [[F^{mn}_1\l^\g]_0 A_\b^2]_0\,,
}
which is not the result displayed in (the last line of) \UoVt. We need to do further processing
to obtain the last line in \UoVt.

Notice that in the second term the SYM superfields
$F^{mn}_1$ and $A_\b^2$ appear in different normal ordered brackets. Therefore
an expression for a massive superfield cannot be identified as the singularity
between $F^{mn}_1$ and $A_\b^2$ has not been taken into account.
However, using the identity \thielid\ followed by \ABCNONO,
the normal ordered bracket from \charac\ builds up and we get
\eqnn\manip
$$\eqalignno{
[[F^{mn}_1\l^\g]_0 A_\b^2]_0 &= [F_1^{mn}[\l^\g A_\b^2]_0]_0
+ [\p\l^\g [F^{mn}_1 A^2_\b]_1]_0 &\manip\cr
&=[\l^\g[F^{mn}_1 A^2_\b]_0]_0 + [\p\l^\g (F^{mn}_1 A^2_\b)]_0\cr
&=[\l^\g(\p F^{mn}_1 A^2_\b)]_0 + [\p\l^\g (F^{mn}_1 A^2_\b)]_0\,,
}$$
and the result is proportional to the massive plane wave $e^{ik_3\cdot X}$.
Similarly,
the first term in \prob\ can be rewritten using \ABCNONO\ and $[\l^\b
N^{mn}]_1=-\half(\g^{mn})^\b{}_\g \l^\g$ as follows
\eqn\Nidtmp{
\ap [\l^\b[N^{mn}(F^{mn}_1 A^2_\b)]_0]_0 = \ap [N^{mn}[\l^\b(F^{mn}_1 A^2_\b)]_0]_0
-{\ap\over2}(\g^{mn})^\b{}_\g[\p\l^\g(F^{mn}_1 A^2_\b)]_0\,.
}
This leads to
\eqnn\finalIf
$$\eqalignno{
I_4 &= \ap [\l^\b[N^{mn}(F^{mn}_1 A^2_\b)]_0]_0
+{\ap\over2}(\g^{mn})^\b{}_\g\Bigl(
[\l^\g(\p F^{mn}_1 A^2_\b)]_0 + [\p\l^\g (F^{mn}_1 A^2_\b)]_0
\Bigr)\cr
&=\ap [N^{mn}[\l^\b(F^{mn}_1 A^2_\b)]_0]_0
+{\ap\over2}(\g^{mn})^\b{}_\g
[\l^\g(\p F^{mn}_1 A^2_\b)]_0\,, &\finalIf
}$$
which is the result we used in the last line of \UoVt.


\newsubsubsec\NOidapp Normal ordering identity

\noindent Using the pure spinor OPEs
\eqn\Nlacont{
\eqalign{
[N^{mn}\l^\a]_1&=\half(\g^{mn}\l)^\a\,,\cr
[\l^\a J]_1 &=-\l^\a\,
}\qquad
\eqalign{
[N^{mn}\l^\a]_{n\ge2}&=0\,,\cr
[\l^\a J]_{n\ge2} &=0\,,
}}
and the identities from the appendix~\opeapp\ we can show
a normal-ordering identity given in \BCpaper\
\eqn\eqtf{
[N^{mn}[\l^\a\l^\b]_0]_0\g^m_{\b\g} =
\half [J[\l^\a\l^\b]_0]_0\g^n_{\b\g}
+{5\over2}\l^\a(\g^n\p\l)_\g
+ \half(\l\g^{mn})^\a(\g^m\p\l)_\g
}
and used in the proof of $QV=0$ in section~\Vmasssec.
To see this note that \ABCNONO\ implies
\eqnn\tmpNlala
$$\eqalignno{
[N^{mn}[\l^\a\l^\b]_0]_0\g^m_{\b\g} &= [\l^\a[N^{mn}\l^\b]_0]_0\g^m_{\b\g}
+[\p[N^{mn}\l^\a]_1\l^\b]_0\g^m_{\b\g}\cr
&=\half [\l^\a[J\l^\b]_0]_0\g^n_{\b\g}+ 2\l^\a(\g^n\p\l)_\g
+ \half(\g^{mn}\p\l)^\a(\l\g^m)_\g \cr
&=\half [J[\l^\a\l^\b]_0]_0\g^n_{\b\g}
-\half\p\l^\a(\g^n\l)_\g + 2\l^\a(\g^n\p\l)_\g + \half(\g^{mn}\p\l)^\a(\l\g^m)_\g\cr
&= \half [J[\l^\a\l^\b]_0]_0\g^n_{\b\g}
+{5\over2}\l^\a(\g^n\p\l)_\g
+ \half(\l\g^{mn})^\a(\g^m\p\l)_\g &\tmpNlala\cr
}$$
where we used \BCpaper\ (to show it, apply $[J,-]_2$ to both sides)
\eqn\Nla{
[N^{mn}\l^\b]_0\g^m_{\b\g}  =
\half[J\l^\b]_0\g^n_{\b\g} +
2(\g^n \p\l)_\g
}
to arrive at the second line while
\ABCNONO\ has been used to arrive at the third line with
$[\l^\a[J\l^\b]_0]_0 =  [J[\l^\a\l^\b]_0]_0 - \p\l^\a\l^\b$.
Finally, $(\p\l\g^m)_\a(\l\g^m)_\b + (\p\l\g^m)_\b(\l\g^m)_\a =0$
leads to the fourth line and the identity \eqtf\ is demonstrated.

\appendix{B}{Equations of motion of massive superfields}
\applab\eomapp

\paragraph{From massless SYM in the OPE gauge}
We will check that the equations of motion for the first-level
massive superfields in our representation given in \Beom\ to \Feom\
are implied by the linearized SYM superfield
equations of motion \SYMBRST.

The massive equations of motion in a BRST language involve the combinations
$(\l B)_\a$, $(\l H_m)$, $(C\l)^\a$ and $(\l F)_{mn}$ defined
in \lfields. Therefore it will be convenient to list these superfields after
contracting the definitions \Balbe\ to \Famn\ with the pure spinor
$\l$:
\eqnn\laBal
\eqnn\lHmsimple
\eqnn\Cla
\eqnn\laF
$$\eqalignno{
(\l B)_\a & =
- 2i\ap k^2_m(\g^m W_1)_\a V_2
	- i\ap  k^1_m(\g_n W_1)_\a (\l\g^{mn}A_2)
	-\frac{\ap }{2}F^{1}_{mn}(\g^{mn}D)_\a V_2 \cr
&=- 2i\ap k^2_m(\g^m W_1)_\a V_2
+ i\ap k_1^m (\g^n W_1)_\a (\l\g^n \g^m A_2)
+ i\ap k_1^m (\l\g^n W_1) (\g^n \g^m A_2)_\a \cr
&\quad{}+{\ap\over2}F_1^{mn}(\l\g^p\g^{mn})_\a A_2^p
+ {\ap\over2}Q\bigl(F^1_{mn}(\g^{mn}A_2)_\a\bigr)&\laBal\cr
(\l H)_m &=
A^{1}_m V_2
+ 2\ap k^1_m(k^2\cdot A^{1}) V_2
-2 i\ap  k^1_m W^\b_1D_\b V_2
- \frac{i\ap }{2}k^1_m F^{1}_{np}(\l\g^{np}A_2)\cr
& =A^m_1 V_2
	+ 2\ap k^1_m(k^2\cdot A^1) V_2
	- 2i\ap k^1_m(\l\g^n W_1)A^2_n
	- 2i\ap k^1_mQ(W_1A_2)\,, \cr
& = -2i\ap\Bigl( k_2^n F_1^{mn}V_2
+ k_1^m (\l\g^n W_1)A_2^n + k_1^m Q(W_1A_2)\Bigr)\,,&\lHmsimple\cr
(C\l)^\a &= W_1^\a V_2 &\Cla\cr
(\l F)_{mn}&=F_1^{mn}V_2 &\laF
}$$
In order to derive the above representations one uses
the gamma matrix identity
$\g^m_{\a(\b}\g^m_{\g\delta)}=0$, the Dirac
equation, the linearized SYM equations of motion \SYMBRST\
as well as the pure spinor constraint. In particular,
\eqn\bex{
-F_1^{mn}(\l\g^{mn}D)V_2 = Q\bigl(F_1^{mn}(\l\g^{mn} A_2)\bigr)\,.
}
In addition, the first massive state condition $-2\ap (k_1\cdot k_2) = 1$
implies that
\eqn\Ffirst{
-2i\ap k_2^n F_1^{mn}V_2 = A_1^m V_2 + 2\ap (k_2\cdot A_1)k_1^m V_2
}
as easily seen after expanding the linearized field-strength $F_1^{mn} = 
ik_1^m A_1^n - ik_1^n A_1^m$.

A straightforward calculation using the usual set of identities
leads to
\eqnn\QlaBal
$$\eqalignno{
Q(\l B)_\a &= (\l\g^m)_\a\bigl[ -2i\ap k_2^n F_1^{mn}V_2
- 2i\ap k_1^m (\l\g^n W_1) A_2^n
- i\ap k_1^n  Q(W_1\g^m\g^n A_2)\bigr] \cr
&= (\l\g^m)_\a (\l H_m) + (\l\g^m)_\a \bigl[
2i\ap k_1^m Q(W_1A_2)
- i\ap k_1^n  Q(W_1\g^m\g^n A_2)\bigr]\cr 
&=(\l\g^m)_\a (\l H_m)\,.&\QlaBal
}$$
To arrive at the last line, note that
the two BRST-exact terms vanish after using $\g^m \g^n = - \g^n \g^m +
2\eta^{nm}$ and the Dirac equation.

Now, taking into account that $\l\g^m W$ is BRST closed
and using the SYM equations of motion \SYMBRST, the rewritten expression
\lHmsimple\ leads to
\eqn\QlH{
Q(\lambda H)_m = -2\ap(k_1\cdot k_2) (\l\g^m W_1)V_2 = (\l\g^m C\l)
}
where we used the first massive state condition $-2\ap(k_1\cdot k_2) = 1$ from
\firstcond\ and the definition \Cla.
This proves the equation of motion \Heom.

The equation of motions \Ceom\ and \Feom\ follow immediately from
the linearized equations \SYMBRST\ and the definitions \Cla\ and \laF.

\newsubsec\Bexaapp $(\l B\l)$ is BRST exact

It is easy to see from \QlaBal\ and the pure spinor constraint that $(\l B\l)$ is BRST closed.
We will now show that it is also BRST exact.

From \laBal\ and the identities $(\l\g^m)_\a(\l\g^m)_\b=0$ and $(\l\g^{mnp}\l)=0$ it follows
that
\eqnn\exproof
$$\eqalignno{
(\l B\l) &=- 2i\ap k^2_m(\l\g^m W_1) V_2
+ {\ap\over2}Q\bigl(F^1_{mn}(\l\g^{mn}A_2)\bigr) &\exproof\cr
&=-2\ap Q\Bigl(i(k_1\cdot A_2)V_1 +A^m_1(\l\g^m W_2)
+i(k_2\cdot A_1)V_2
-{1\over4}F^1_{mn}(\l\g^{mn}A_2)\Bigr)\,,
}$$
where we used  the linearized equations \SYMBRST. Therefore $\l^\a\l^\b(B_{\a\b}-D_\a\Lambda_\b)=0$
with
\eqn\Lamdef{
\Lambda_\b = -2\ap \Bigl(i(k_1\cdot A_2)A^1_\b +A^m_1(\g^m W_2)_\b
+i(k_2\cdot A_1)A^2_\b
-{1\over4}F^1_{mn}(\g^{mn}A_2)_\b\Bigr)\,.
}


\listrefs

\bye